\documentclass[aps,prb,twocolumn,floatfix]{revtex4-1}
\usepackage{times}
\usepackage{epsfig}
\usepackage{amsmath}
\usepackage{amssymb}
\usepackage{graphicx}
\usepackage{dsfont}
\usepackage[colorlinks=true,citecolor=red,linkcolor=blue]{hyperref}

\newcommand{\angstrom}{\mbox{\normalfont\AA}}

\newcommand{\qbc}{\mathrm{QBC}}
\newcommand{\bfk}{\mathbf{k}}
\newcommand{\bfa}{\mathbf{a}}

\newcommand{\bfA}{\mathbf{A}}
\newcommand{\bfR}{\mathbf{R}}
\newcommand{\bfK}{\mathbf{K}}
\newcommand{\tldA}{\tilde{\mathbf{A}}}
\newcommand{\bfd}{\mathbf{\delta}}
\newcommand{\tph}{\tilde{\phi}}
\newcommand{\tpeff}{t_{\mathrm{eff}}'}
\newcommand{\tseff}{t_{\mathrm{eff}}^{\sigma}}
\newcommand{\tlda}{\tilde{a}}

\begin{document}

\title{Floquet states in (LaNiO$_3$)$_2$/(LaAlO$_3$)$_N$ heterostructures grown along the (111) direction}
\author{Liang Du}
\affiliation{Department of Physics, The University of Texas at Austin, Austin, TX 78712, USA}
\author{Gregory A. Fiete}
\affiliation{Department of Physics, The University of Texas at Austin, Austin, TX 78712, USA}

\begin{abstract}
Using Floquet-Bloch theory we study the effect of circularly and linearly polarized light on the electronic structure of 
(LaNiO$_3$)$_2$/(LaAlO$_3$)$_N$ heterostructure grown along the (111) direction.  In equilibrium, a tight-binding fit to the
first principles band structure shows that nearest-neighbor hopping plays a dominant role while second-neighbor hopping breaks the particle-hole symmetry and determines the finer band features.  The four bands of the LaNiO$_3$ bilayer exhibit both quadratic band touching points and Dirac points. By varying the amplitude of the incident light, one can independently tune the first and second-neighbor hopping for fixed frequency, which leads to considerable control over the Floquet band structure.  We investigate this control in detail, and study how the quadratic and Dirac band touchings are influenced by the polarization and intensity of the light. We derive effective 2-band Hamiltonians (for both quadratic and Dirac band touching points) that accurately captures these results. We further study an extended model which explicitly includes oxygen $p$-orbitals and compare the results to the effective model that contains only the nickel $d$-orbitals. We conclude with a computation of the frequency dependent optical Hall conductivity using the full four band model and analyze the various inter-band contributions  of the Floquet modes. 
\end{abstract}

\date{\today}
%\pacs{73.43.-f, 78.20.Bh, 03.65.Vf}
\maketitle

\section{INTRODUCTION}
\label{sec:intro}
% 0. realistic materials
% 1. multi-band effect of previous studies
% 2. effect of second nearest neighbor hopping terms in equilibrium 
% conclude: answer two questions :  (1) and (2)
% Previous studies would like to include only nearest neighbor hopping terms.
Topological insulators have been one of the most active research topics in condensed matter physics over the past decade with dramatic advances in both theoretical and experimental areas.\cite{Moore:nature2010, Hasan:rmp2010, Qi:rmp2011, Ando:jpsj2013}  Correlated topological phases in particular are expected to be exceptionally rich in their phenomenology,\cite{Maciejko:np15,Stern:arcmp16,Amaricci:prl15,Amaricci:prb16} with transition metal oxides\cite{Pesin:np10,Krempa:arcm14,Witczak:prb12,Maciejko:prl14,Kargarian:prl13,Ruegg:prl12,Shitade:prl09,Yang:prl14,Rau:arcm16} and SmB$_6$ receiving considerable attention.\cite{Dzero:prl10,Neupane:nc13,Zhang:prx13,Frantzeskakis:prx13,Jiang:nc13}
Heterostructure of transition metal oxides involving partially filled $d$-bands have been one candidate for a correlated 
topological insulator, particularly the quantum anomalous Hall effect.\cite{Fiete:jap2015,Xiao:natcom2011,Hwang:sr16,Hu:sr15,Wang:prb15} 

One specific example that has been theoretical proposed in the literature\cite{Ruegg_Top:prb13,Ruegg:prb2011,Ruegg:prb12,Yang:prb11a} and experimentally studied\cite{Middey:apl12} is the (LaNiO$_3$)$_2$/(LaAlO$_3$)$_N$ system grown along the (111) crystalline axis. 
Density functional theory calculations\cite{Ruegg_Top:prb13,Ruegg:prb12} suggest that among the four bands closest to the Fermi energy, the upper and lower are nearly flat and quadratic touching points connect these to intermediate energy bands that cross each other at two inequivalent Dirac points located at  the corners of the hexagonal first Brillouin zone.\cite{Ruegg_Top:prb13,Ruegg:prb12}  A tight-binding fit to the first principles band structure shows that the nearest neighbor hopping terms dominate the band structure, while the next-nearest neighbor hopping terms break the particle-hole symmetry and determine the finer details of the band structure.\cite{Ruegg_Top:prb13,Ruegg:prb12} Prior theoretical studies\cite{Ruegg:prb2011, Xiao:natcom2011, Yang:prb2011} on this system showed that purely local (on-site) Coulomb interactions can induce a topological insulator (quantum anomalous Hall) phase, even if only the nearest-neighbor hopping in the generalized tight-binding model are retained.\cite{Ruegg:prb2011, Yang:prb2011}  Including the second-neighbor hopping term has a very small quantitative (and no qualitative) effect on the phase diagram of the system.\cite{Ruegg_Top:prb13,Ruegg:prb12} However, as we report in this work, the presence of both first and second-neighbor hopping have a dramatically greater consequence if the system is driven out of equilibrium by an applied laser field.

In parallel with with the theoretical study of correlated topological insulators has been a substantial effort directed at engineering topological band structures in a non-topological system through the application of a periodic drive,\cite{Kitagawa:prb10,Rudner:prx13,Katan:prl13,Lindner:prb13,Dora:prl12,Inoue:prl10,Cayssol:pss13,Kitagawa:prb11,Iadecola:prl13,Ezawa:prl13,Kemper:prb13,Rechtsman:nat13} particularly that originating in a laser field.  In the solid state context, some experimental progress has been made in this direction.\cite{Fregoso:prb13,Wang:sci13,Mahmood:np16} Such Floquet topological insulators have considerable new physics associated with them compared to their equilibrium topological counterparts:  The bulk-boundary correspondence breaks down,\cite{Farrell:prb16,Torres:prl14,Rudner:prx13,Kitagawa:prb11} interaction effects are more complicated to understand,\cite{Seetharam:prx15,Grushin:prl14,Polkovnikov:rmp11} and even the steady-state occupation of the Floquet bands can be subtle.\cite{Seetharam:prx15,Iadecola:prb15,Iadecola_2:prb15,Hossein:prb16} 

Previous studies of topological Floquet systems have mainly focused on the Dirac points in two-dimensional 2-band systems related to the A-B sub-lattice in graphene systems.\cite{Inoue:prl10,Oka:prb09,Kitagawa:prb11,Iadecola:prl13,Ezawa:prl13,Gu:prl11,Hossein:prb14,Aditiprb91-2015,Aditiprb92-2015,Hossein:prb16} The generation of a Floquet-Bloch band structure with a laser can be understood as an optical dressing of the original electronic band structure.\cite{Rudner:prx13}  In particular, a dynamical band gap can be opened by virtual one-photon absorption and emission processes.\cite{Oka:prb2009, Kitagawa:prb2011}  Recently these studies were extended to include more than two bands, as well as quadratic band touching points in the band structure, through a model on the kagome lattice.\cite{Du:prb2017}  In Ref.[\onlinecite{Du:prb2017}] it was found that the multi-band nature did not influence the physics of the gap opening at the Dirac points, but new physics emerged at the quadratic band touching point. The main result is that the quadratic band touching point has a gap opened by virtual two-photon (as opposed to one-photon for a Dirac point) absorption and emission processes.\cite{Du:prb2017} For the kagome lattice, it is possible to derive an effective 2-band model of the quadratic band touching point \cite{Sun:prl09} that captures the general features of the gap opening.  

In this work, we focus on the (111) bilayer LaNiO$_3$, for which there are both experimental realizations\cite{Middey:apl12} and theoretical calculations of the band structure.\cite{Ruegg_Top:prb13,Ruegg:prb12}  We would like to understand the answer to the following two questions: 
(1) Can small further-neighbor (beyond first-neighbor) hopping terms have a more dramatic effect in a Floquet system than in equilibrium? 
(2) Are there any important effects in the Floquet band structure as the total number of equilibrium bands is increased?
We answer both of these questions in the affirmative, and provide the details in the main sections of this manuscript.

Our paper is organized as follows. In Sec.\ref{sec:model}, we describe the generalized tight-binding model used to quantitatively fit the first principles calculations of the (LaNiO$_3$)$_2$/(LaAlO$_3$)$_N$ heterostructure. In Sec.\ref{sec:drive} and Sec.\ref{sec:floquet} we describe the Floquet-Bloch bands resulting from both linearly and circularly polarized laser fields. We derive an effective 2-band Hamiltonian to describe the behavior around the quadratic touching and the Dirac points in Sec.\ref{sec:lowenergy}.
Then in Sec.\ref{sec:optical}, we compute the finite frequency optical conductivity of the material for different laser parameters.  Finally, in Sec.\ref{sec:discussion} we summarize the main conclusions of this work.

\section{Model and method}
\label{sec:model}
\begin{figure}[h]
\includegraphics[width=1.0\linewidth]{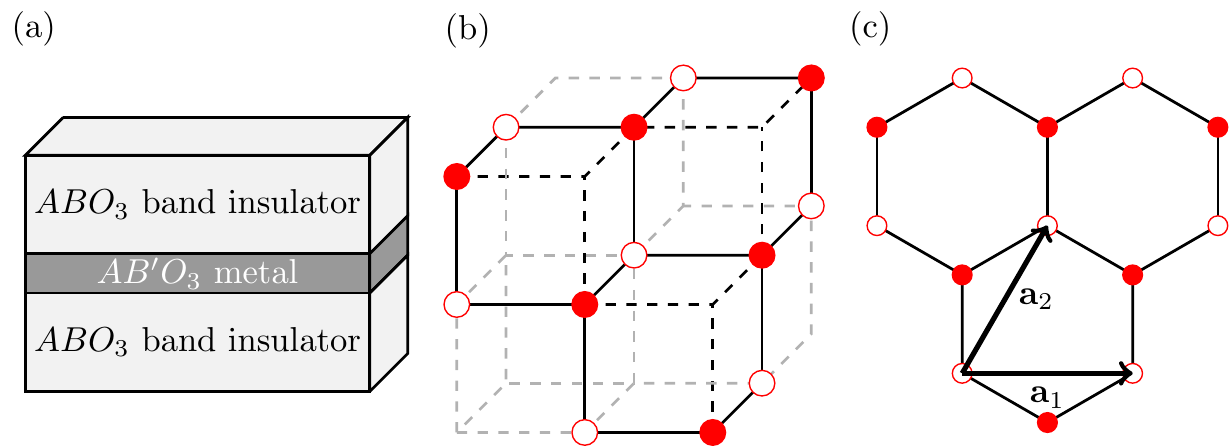}  %
\caption{(Color online) 
(a) Transition metal oxide heterostructure grown along (111) direction of the form AB$'$O$_3$/ABO$_3$/AB$'$O$_3$. 
    The shaded area consists of the (111) bilayer LaNiO$_3$, and the light area consists of the non-magnetic band insulator LaAlO$_3$. 
(b) The locations of the transition metal ions Ni in (111) bilayers of perovskite structured LaNiO$_3$ are shown. Filled (open) 
circles represent ions in top (bottom) layer. The lattice constant is $a_0= 3.82 \angstrom$.
(c) Buckled honeycomb lattice formed in the (111) bilayer LaNiO$_3$. The lattice constant is $\tilde{a} = \sqrt{2/3}a_0 =3.12 \angstrom$. 
The primitive lattice vectors are chosen as $\bfa_1=(\sqrt{3}, 0)\tilde{a}$, $\bfa_2=(\sqrt{3}/2,3/2)\tilde{a}$. 
For convenience, an additional vector is defined as $\bfa_3=\bfa_2-\bfa_1$.
}
\label{fig:honeycomb_lattice}
\end{figure}

We consider the thin film oxide geometry shown in Fig.\ref{fig:honeycomb_lattice}(a), a bilayer of metallic LaNiO$_3$ sandwiched between non-magnetic band insulator LaAlO$_3$ layers. The Fermi level lies in the $e_g$ bands derived from the Ni states, and we focus on those orbitals in our model below.  The lattice structure formed by the transition metal ions Ni are shown in 
Fig.\ref{fig:honeycomb_lattice}(b).  The distance between two nearest-neighbors is $a_0=3.82 \angstrom$.\cite{Ruegg:prb12} 
In the bilayer geometry, the transition metal ions form a buckled honeycomb lattice consisting of two trigonal layers, as shown in Fig.\ref{fig:honeycomb_lattice}(c). 

In the following, we use $A$ to denote ions in the top layer and $B$ the bottom layer. 
The lattice constant in the buckled honeycomb lattice (projected into the plane) is $\tlda = \sqrt{2/3} a_0=3.12\angstrom$. 
We choose the primitive lattice vector as $\bfa_1=(\sqrt{3}, 0)\tlda$ and $\bfa_2=(\sqrt{3}/2, 3/2)\tlda$.
For notational convenience, an additional vector $\bfa_3$ is defined as $\bfa_3=\bfa_2-\bfa_1$. We also defined 
three nearest neighbor vectors,
$\delta_x=(-\sqrt{3}/2,-1/2)\tlda, \delta_y=(\sqrt{3}/2, -1/2)\tlda, \delta_z=(0,1)\tlda$.
The reciprocal lattice vectors are $b_1=(\sqrt{3},-1)2\pi/3\tlda$ and $b_2=(0, 2)2\pi/3\tlda$. The first Brillouin zone is a 
hexagon with $K'=(-4\pi/3\sqrt{3}\tlda, 0)$ and $K=(4\pi/3\sqrt{3}\tlda, 0)$ located on the $x$-axis and $M_\mp=(0, \mp2\pi/3\tlda)$ on the $y$-axis.
In order to clearly exhibit the effect of the laser on the quadratic touching and Dirac points, 
the Floquet-Bloch electronic band structures are along the path $K'-\Gamma-K$ or $M_--\Gamma-M_+$.

The generalized tight-binding model for the (111) bilayer film can be expressed as,\cite{Ruegg:prb12,Fiete:jap2015} 
\begin{align}
     H   = &\sum_{\bfR\in A} \sum_{u} [d^\dagger(\bfR) t_u d(\bfR+\bfd_u) + h.c.] \nonumber\\
         + &\sum_{\bfR\in A} \sum_{u} [d^\dagger(\bfR) t_{u,u+1} d(\bfR+\bfd_u-\delta_{u+1}) + h.c.] \nonumber\\
         + &\sum_{\bfR\in B} \sum_{u} [d^\dagger(\bfR) t_{u,u+1} d(\bfR-\bfd_u+\delta_{u+1}) + h.c.]
         \label{eq:ni-ni-tb}
\end{align}
where $d^\dagger=(d_{3z^2-r^2}^\dagger, d_{x^2-y^2}^\dagger)$, $u=x,y,z$, and $u+1=y,z,x$, with those particular orderings, written out in Eq.\eqref{eq:next-nearest-neighbour}.
The transfer matrix between $e_g$ orbitals in different direction are determined by 
using the standard Slater-Koster procedure. The nearest neighbor (NN) hopping matrices are,
\begin{eqnarray}
\label{eq:nearest-neighbour}
     t_z &=& - 
     \begin{pmatrix}
         t_\sigma & 0 \\
         0 & t_\delta
     \end{pmatrix},  \nonumber\\
     t_x &=& - \frac{1}{4}
     \begin{pmatrix}
         t_\sigma+3t_\delta & \sqrt{3}(t_\delta-t_\sigma) \\
         \sqrt{3}(t_\delta-t_\sigma) & 3t_\sigma + t_\delta
     \end{pmatrix}, \nonumber\\
     t_y &=& - \frac{1}{4}
     \begin{pmatrix}
         t_\sigma+3t_\delta & \sqrt{3}(t_\sigma - t_\delta) \\
         \sqrt{3}(t_\sigma-t_\delta) & 3t_\sigma + t_\delta
     \end{pmatrix}.
\end{eqnarray}
The next-nearest neighbor (NNN) hopping matrices are,
\begin{eqnarray}
\label{eq:next-nearest-neighbour}
     t_{xy} &=& 
     \begin{pmatrix}
         -t'/2 & 0 \\
         0   & 3t'/2
     \end{pmatrix},  \nonumber\\
     t_{yz} &=&  
     \begin{pmatrix}
         t' & \sqrt{3}t'/2 \\
         \sqrt{3}t'/2 & 0
     \end{pmatrix}, \nonumber\\
     t_{zx} &=& 
     \begin{pmatrix}
         t' & -\sqrt{3}t'/2 \\
         -\sqrt{3}t'/2 & 0
     \end{pmatrix}.
\end{eqnarray}

In momentum space, the tight-binding Hamiltonian on the honeycomb lattice takes the form,
\begin{align}
\label{eq:htbk-AB}
     H(\bfk) = 
     \begin{pmatrix}
          H_{AA}(\bfk) & H_{AB}(\bfk) \\
          H_{BA}(\bfk) & H_{BB}(\bfk)
     \end{pmatrix}, 
\end{align}
with 
\begin{align}
    &H_{AA}(\bfk) = 2 (t_{xy} \cos k_1 + t_{zx} \cos k_2 + t_{yz} \cos k_3) = H_{BB} \nonumber\\
    &H_{AB}(\bfk) = t_{x} e^{- i k_2} + t_{y} e^{-i k_3} + t_{z} = H_{BA}^\dagger,
\end{align}
where we used $k_i = \bfk \cdot \bfa_i$, and $A,B$ denote sub-lattice in the buckled honeycomb lattice. 
By substituting the NN hopping matrix Eq.(\ref{eq:nearest-neighbour}) and NNN hopping matrix Eq.(\ref{eq:next-nearest-neighbour}) into Eq.(\ref{eq:htbk-AB}), we write
the tight-binding Hamiltonain explicitly as, 
\begin{equation}
\label{eq:htbk}
     H(\bfk) = 
     \begin{pmatrix}
     \tilde{\epsilon}_{aa\bfk}  & \tilde{\epsilon}_{ab\bfk}  &       {\epsilon}_{aa\bfk} &       {\epsilon}_{ab\bfk} \\
     \tilde{\epsilon}_{ab\bfk}  & \tilde{\epsilon}_{bb\bfk}  &       {\epsilon}_{ab\bfk} &       {\epsilon}_{bb\bfk} \\
           {\epsilon}_{aa\bfk}^*&       {\epsilon}_{ab\bfk}^*& \tilde{\epsilon}_{aa\bfk} & \tilde{\epsilon}_{ab\bfk} \\
           {\epsilon}_{ab\bfk}^*&       {\epsilon}_{bb\bfk}^*& \tilde{\epsilon}_{ab\bfk} & \tilde{\epsilon}_{bb\bfk}
     \end{pmatrix},
\end{equation}
with the matrix elements given by
\begin{align}
     & \epsilon_{aa\bfk} = -t_\sigma - \frac{1}{2}(t_\sigma+3t_\delta) \cos(\sqrt{3}k_x/2)e^{-i3k_y/2}, \nonumber \\
     & \epsilon_{bb\bfk} = -t_\delta - \frac{1}{2}(3t_\sigma+t_\delta) \cos(\sqrt{3}k_x/2)e^{-i3k_y/2}, \nonumber \\
     & \epsilon_{ab\bfk} = -i\frac{\sqrt{3}}{2}(t_\sigma-t_\delta) \sin(\sqrt{3}k_x/2)e^{-i3k_y/2}, \nonumber \\
     & \tilde{\epsilon}_{aa\bfk} = t' [4 \cos(\sqrt{3}k_x/2) \cos(3k_y/2) - \cos(\sqrt{3}k_x)], \nonumber \\
     & \tilde{\epsilon}_{bb\bfk} = 3t' \cos(\sqrt{3}k_x), \nonumber \\
     & \tilde{\epsilon}_{ab\bfk} = 2\sqrt{3} t' \sin(\sqrt{3}k_x/2) \sin(3k_y/2),
\end{align}
where we used $a,b$ to denote the two $e_g$ orbitals with $|a\rangle = |d_{3z^2-r^2}\rangle$, $|b\rangle=|d_{x^2-y^2}\rangle$.

By fitting the local density approximation (LDA) band structure with tight-binding model parameters, previous studies 
show the dominant effect is from nearest neighbor hopping $t_\sigma\approx 0.6$eV. 
The next biggest contribution is from next-nearest neighbor hopping $t' \approx 0.1t_\sigma$. \cite{Ruegg:prb12} 
The direct overlap integral $t_\delta$ is vanishing small. Previous equilibrium studies (absence of laser) show that the next nearest neighbor hopping breaks the particle-hole symmetry of band structure. However, its effect 
on the quadratic touching and Dirac points is negligible. In this work, we show that the next-nearest 
neighbor hopping terms play an important role in the non-equilibrium (laser-driven) case. 

However, in order to describe the (111) LaNiO$_3$ bilayer more quantitatively, we fix $t_\sigma=0.6$eV and 
set $t'/t_{\sigma}$ as an adjustable parameter in the range $(0.0,0.1)$. The electronic band structure with $t'/t_\sigma=0.0$ and 
$t'/t_\sigma=0.1$ of Eq.\eqref{eq:htbk} are demonstrated in Fig.\ref{fig:floquet-band-nn-lin}(a) and Fig.\ref{fig:floquet-band-nnn-lin}(a-b), respectively. 
We find band 1 and 2  (3 and 4) touch at the $\Gamma$ point, resulting in two quadratic touching points. Band 2 and 3 touch at two inequivalent Dirac points, $K'$ and $K$. 
We have labeled the above electronic bands in its energy ascending order.  We now turn to the case of an incident laser field. 

\section{Periodic Drive under a laser field}
\label{sec:drive}
When the system is exposed to a normally incident [along the (111) direction] laser field, the canonical momentum of the electron 
is modified through the minimal substitution, $\mathbf{k} \rightarrow \mathbf{k} + \tldA(t)$, where $\tldA(t) = e\mathbf{A}(t)/\hbar$ 
with $\mathbf{A}(t)$ the in-plane laser vector potential, $e$ the electron charge and $\hbar$ the reduced Planck's constant. The Hamiltonian becomes time-dependent,
%>>> time dependent tight-bing hamiltonian in momentum space
    \begin{equation}\label{eq:htbk-t}
        H(\bfk, t) = 
        \begin{pmatrix}
        {H}_{AA}(\bfk, t) &  {H}_{AB}(\bfk, t) \\
        {H}_{BA}(\bfk, t) &  {H}_{BB}(\bfk, t)
        \end{pmatrix},
    \end{equation}
where each element is a $2\times2$ sub-matrix because there exist two active orbitals ($e_g$ orbitals) per site.
Further we need to be careful about $H_{AB}(\bfk, t)$ because we already used a gauge transformation\cite{Aditiprb91-2015}
$d_{kB} \rightarrow d_{k B} e^{i\bfk\cdot\delta_z}$, 
\begin{align}
H_{AB}(\bfk,t) = t_{x} e^{- i \bfk_2 - i\tldA\cdot\delta_x} + t_{y} e^{- i \bfk_3 - i\tldA\cdot\delta_y}
               + t_{z} e^{- i\tldA\cdot\delta_z} \nonumber.
\end{align}
In Eq.\eqref{eq:htbk-t}, the Coulomb gauge is adopted by setting the scalar potential $\phi=0$ and the tiny effect of the direct (Zeeman) coupling of the magnetic field to the spin of the electron is ignored.

Throughout this paper, we use the vector potential $\bfA(t)=A_0[\cos(\Omega t), -\sin(\Omega t)]$ to represent circularly polarized laser fields and 
$\bfA(t) = A_0 \cos(\Omega t)[\cos \theta, \sin\theta]$ to represent linearly polarized laser fields, where $A_0$ and $\Omega$ are the amplitude and frequency of the laser, respectively.
For notational convenience, we define $\tilde{A}_0=eA_0/\hbar$ with unit $1/\tlda$. 

\section{FLOQUET THEORY}
\label{sec:floquet}
A Hamiltonian with a periodic time dependence can be described by Floquet theory,\cite{Rudner:prx13} which states that the solutions to the time-dependent Schr\"{o}dinger's equation (in momentum space) can be expressed as,
\begin{equation}
    |\Psi_{k\alpha}(t)\rangle = e^{-i\epsilon_{k\alpha}t}|\phi_{k\alpha}(t) \rangle 
    = e^{-i\epsilon_{k\alpha}t}\sum_m e^{im\Omega t} |\tph_{k\alpha}^m\rangle,
\end{equation}
with $m=0, \pm 1, \pm 2, \cdots$ and $|\tph_{k\alpha}^m\rangle$ a four component vector (from the two orbitals/site and two sites in the bilayer unit cell) indexed by $\alpha$ which obeys,
\begin{equation}
      \sum_m (H_{nm} + m\hbar\Omega \delta_{nm}) |\tph_{k\alpha}^m\rangle = \epsilon_{k\alpha}|\tph_{k\alpha}^m\rangle,
\end{equation}
with matrix elements of the Floquet Hamiltonian written as,
\begin{align}
    H_{nm}(\bfk) &= \frac{1}{T}\int_{0}^{T} dt e^{-i(n-m)\Omega t} H(k, t) \nonumber\\
              &= 
     \begin{pmatrix}
      H^{AA}_{nm}(\bfk) & H^{AB}_{nm}(\bfk) \\
      H^{BA}_{nm}(\bfk) & H^{BB}_{nm}(\bfk) \\
     \end{pmatrix}.
\end{align}
Here $m$ and $n$ are integers ranging from $-\infty$ to $\infty$.
Thus, the Floquet matrix is an infinite dimensional time-independent matrix. 

In this paper, we consider the laser frequency ($\hbar\Omega$) to be 
in the region smaller (resonant regime) and larger (off-resonant regime) than the bandwidth of the system, respectively.
By shining with off-resonant light, the original electronic bands are renormalized through the virtual photon absorption and emission processes. As a result, The Hamiltonian out of equilibrium can be effectively described by a static Hamiltonian with virtual photon transition (Floquet-Magus expansion). Technically, this is an easier limit to consider, and the interpretation of experimental results will be simpler since they can be understood in the context of an expansion. The large frequency limit avoids resonant transitions and allows an inverse frequency expansion of the Hamiltonian.\cite{Bukov:ap15} In the off-resonant regime, a
truncation of the Floquet components to be in $m,n=-2,-1,0,1,2$ is a good approximation.  

% figures with only nn hopping terms
\begin{figure}[t]
\begin{center}
\includegraphics[width=4cm]{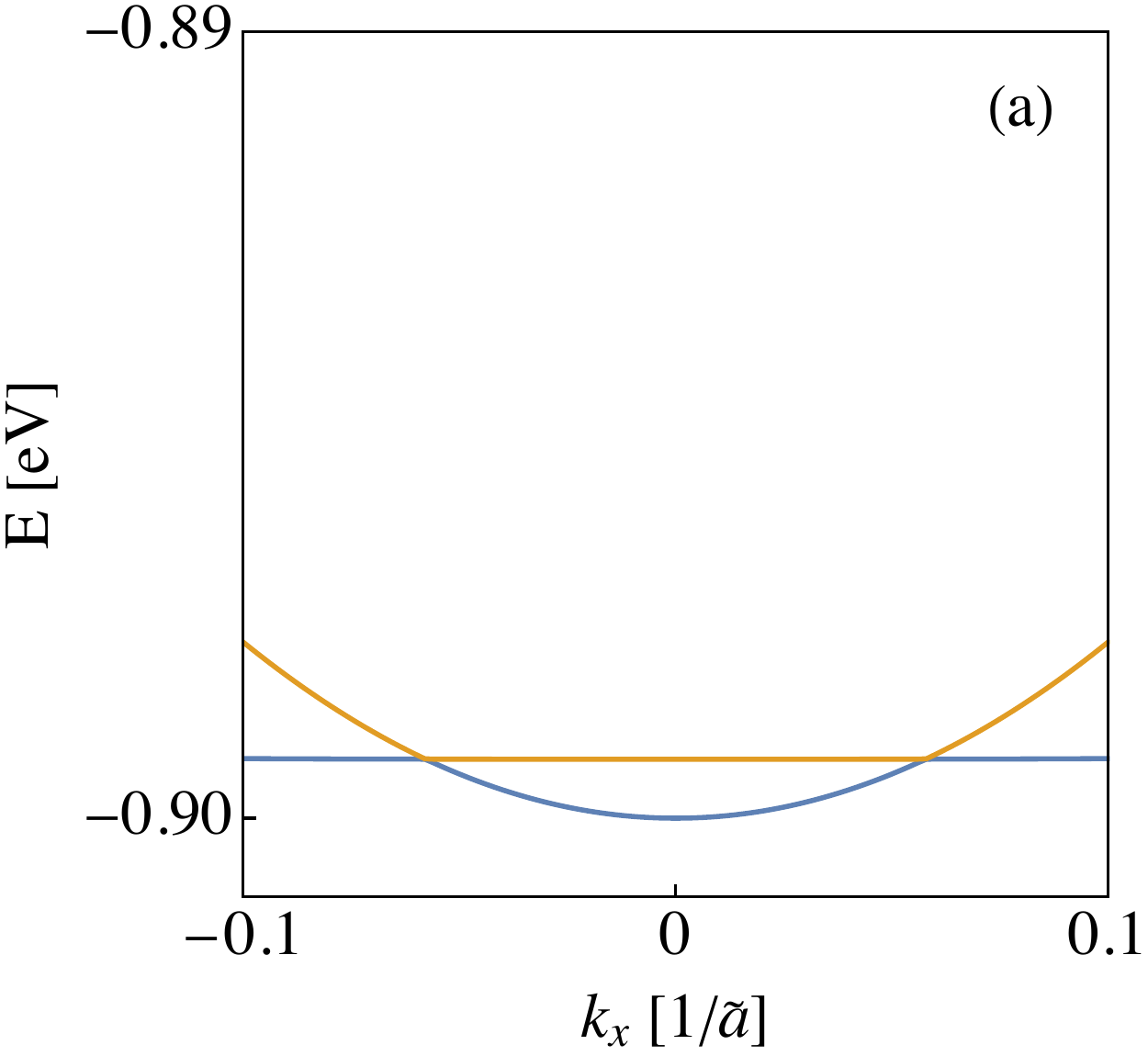}  %
\includegraphics[width=4cm]{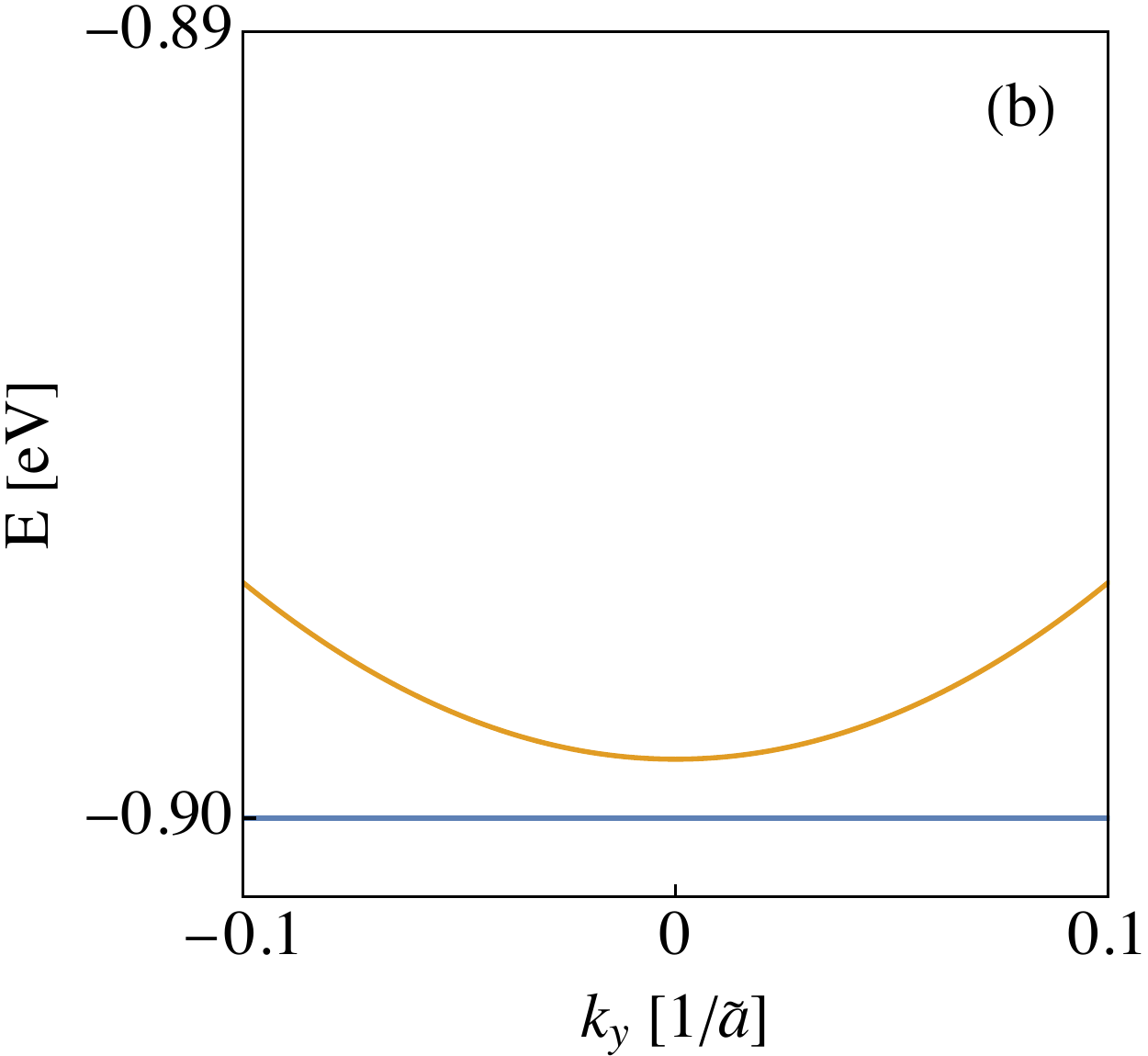}  %
\includegraphics[width=4cm]{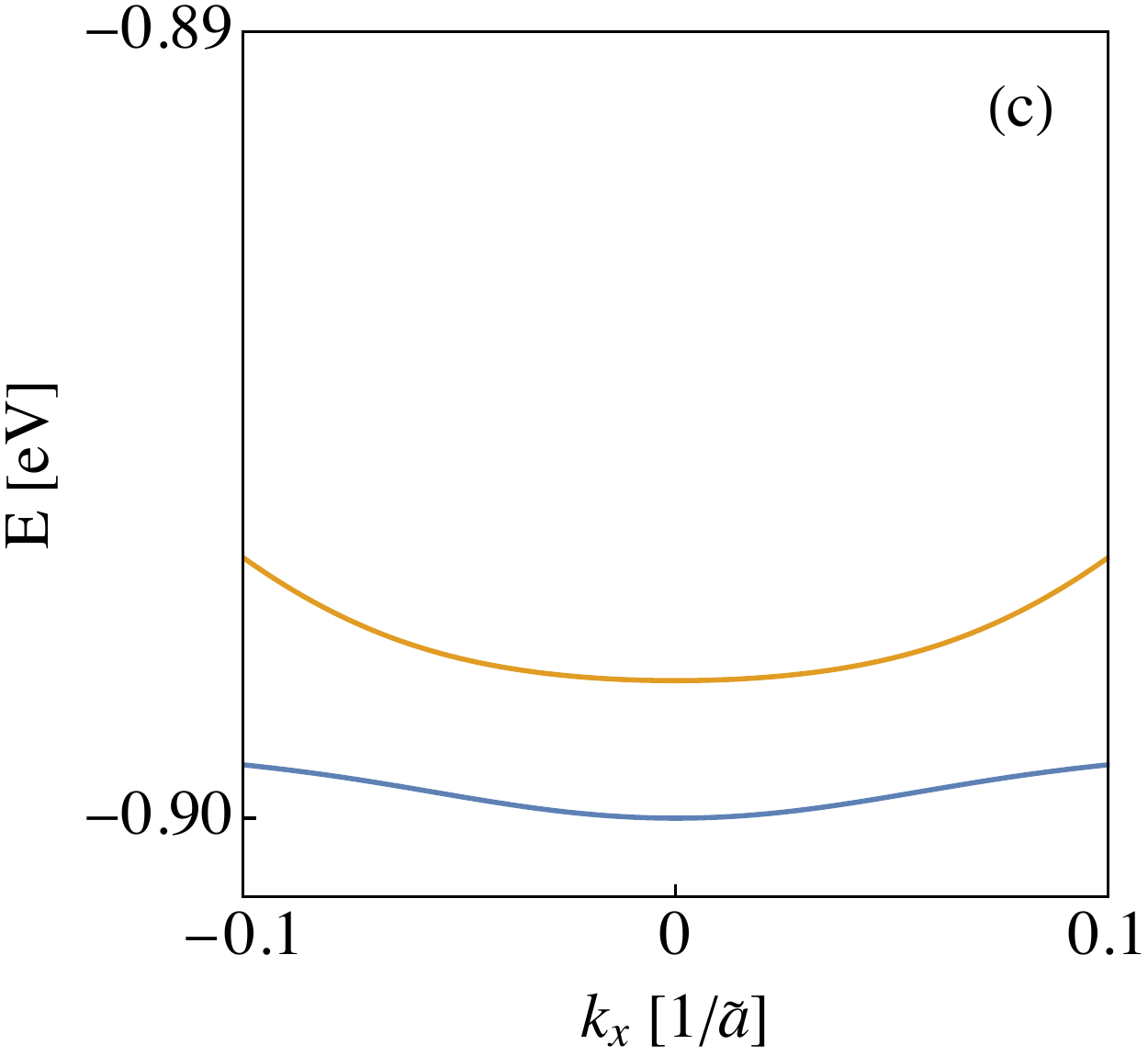}  %
\includegraphics[width=4cm]{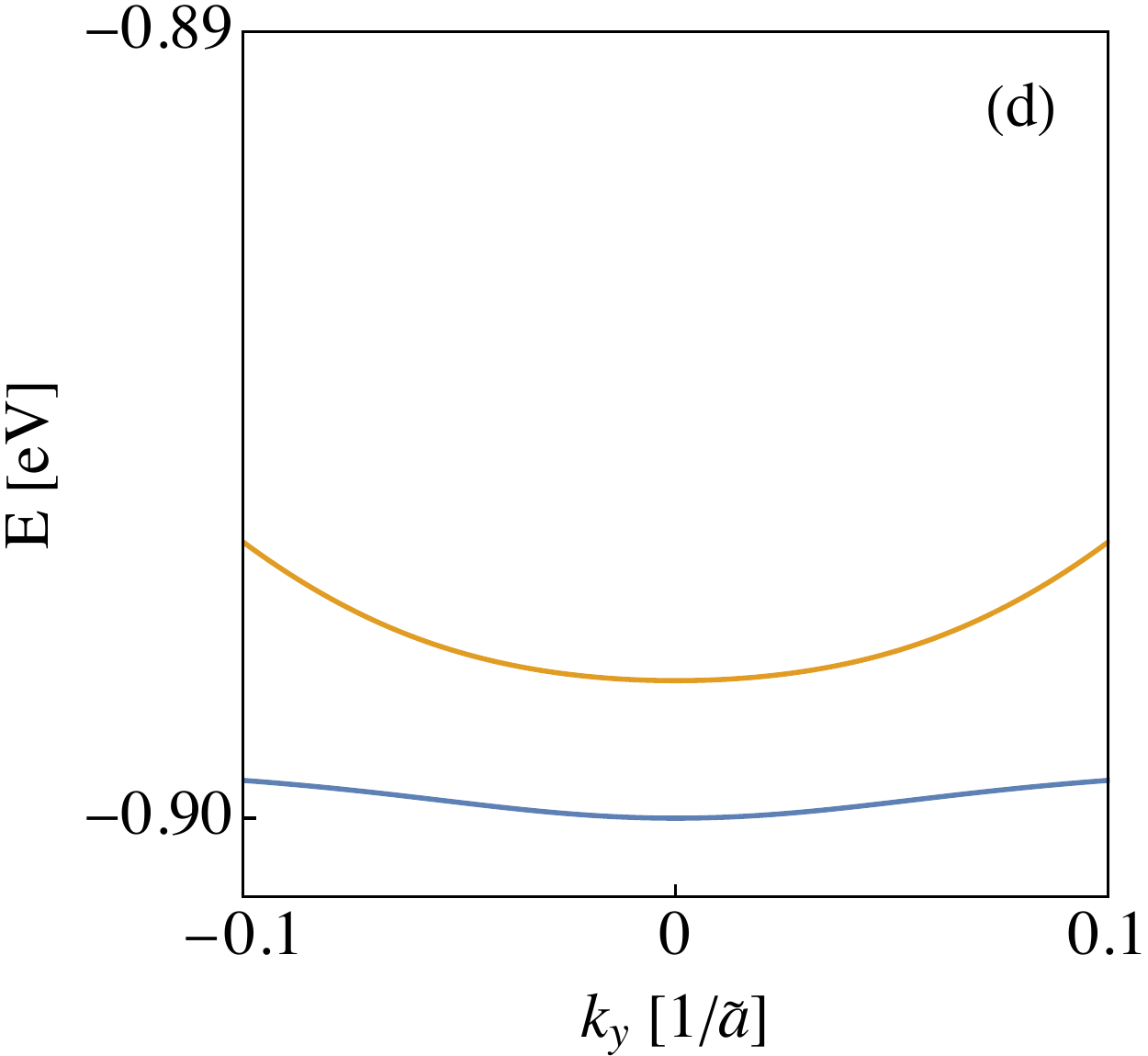}  %
\end{center}
\caption{(Color online) The Floquet-Bloch band structure of the (111) bilayer LaNiO$_3$ embedded in a normally incident linearly polarized light 
$\bfA(t) = A_0\cos(\Omega t)(\cos\theta, \sin\theta)$ with 
$\tilde{A}_0\tlda =\sqrt{2}/10$ and $\hbar\Omega=20.6783$meV (frequency $\nu=\Omega/2\pi=5$THz and intensity $I=116.598$ mW/$\mu$m$^2$). This is a low-energy resonant regime of the drive. Only the dominant nearest neighbor hopping terms are kept in the tight-binding Hamiltonian Eq.\eqref{eq:htbk-t} 
with $t_\sigma=0.6$eV, $t'=0.0$eV. 
(a) $\theta=0$, bands along $k_x (k_y=0)$.
(b) $\theta=0$, bands along $k_y (k_x=0)$.
(c) $\theta=\pi/2$, bands along $k_x (k_y=0)$.
(d) $\theta=\pi/2$, bands along $k_y (k_x=0)$.
}
\label{fig:floquet-band-nn-lin-THz}
\end{figure}

Resonant light will generally lead to much more complicated results, because the precise details of the band structure and associated resonant transitions will influence how the states are modified by the light. This is the regime of heavily overlapping Floquet copies. However, if we are only interested in the region very close to a specific $k$ point, we can still study the Floquet-Bloch band structure around that point in a restricted region of momentum space and analyze the behavior using the low-energy Hamiltonian.\cite{Du:prb2017} In the resonant regime, a
truncation of the Floquet components $m,n$ needs to be tested until convergence is achieved.
% figures with only nn hopping terms
\begin{figure}[t]
\begin{center}
\includegraphics[width=4cm]{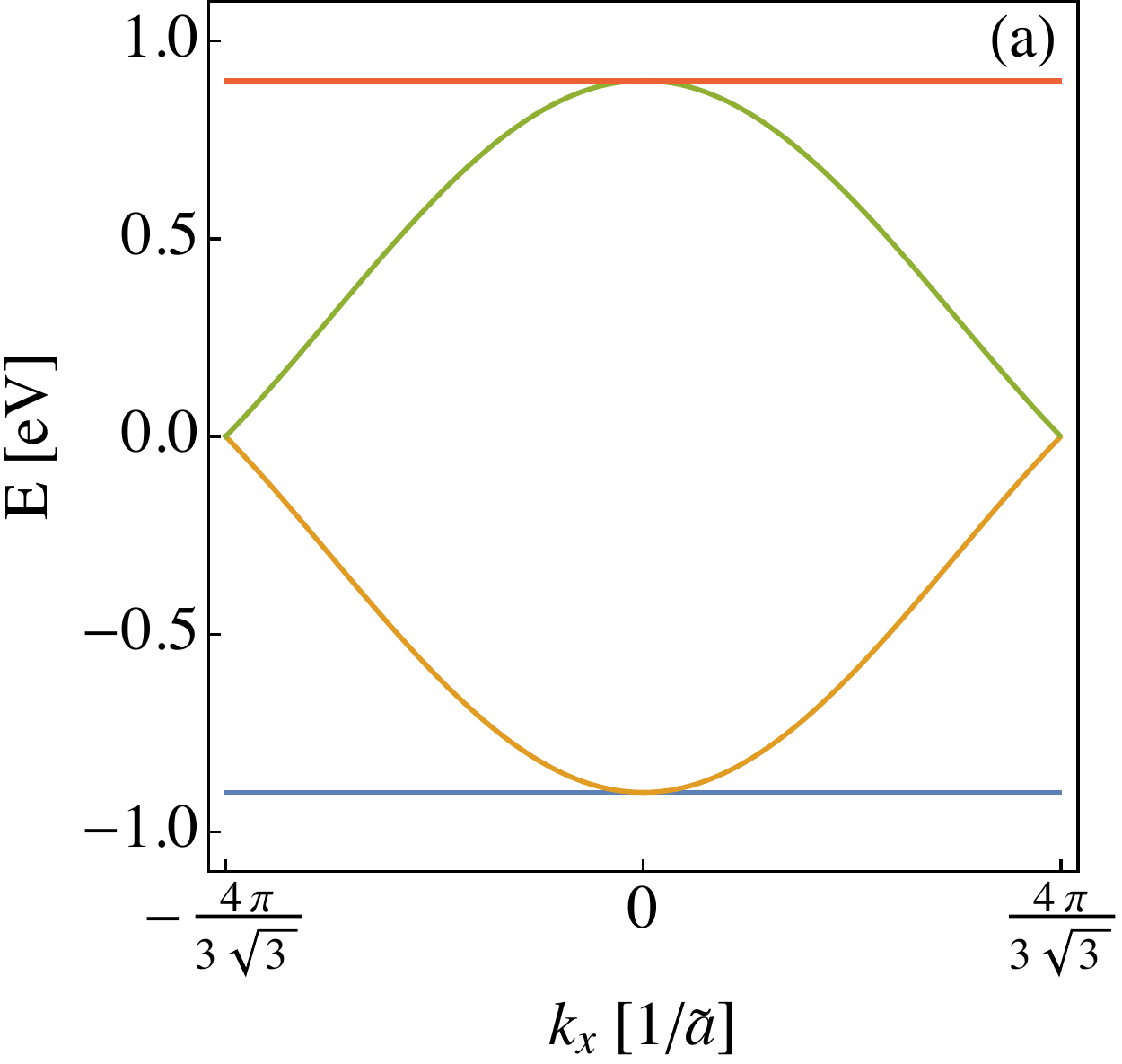}  %
\includegraphics[width=4cm]{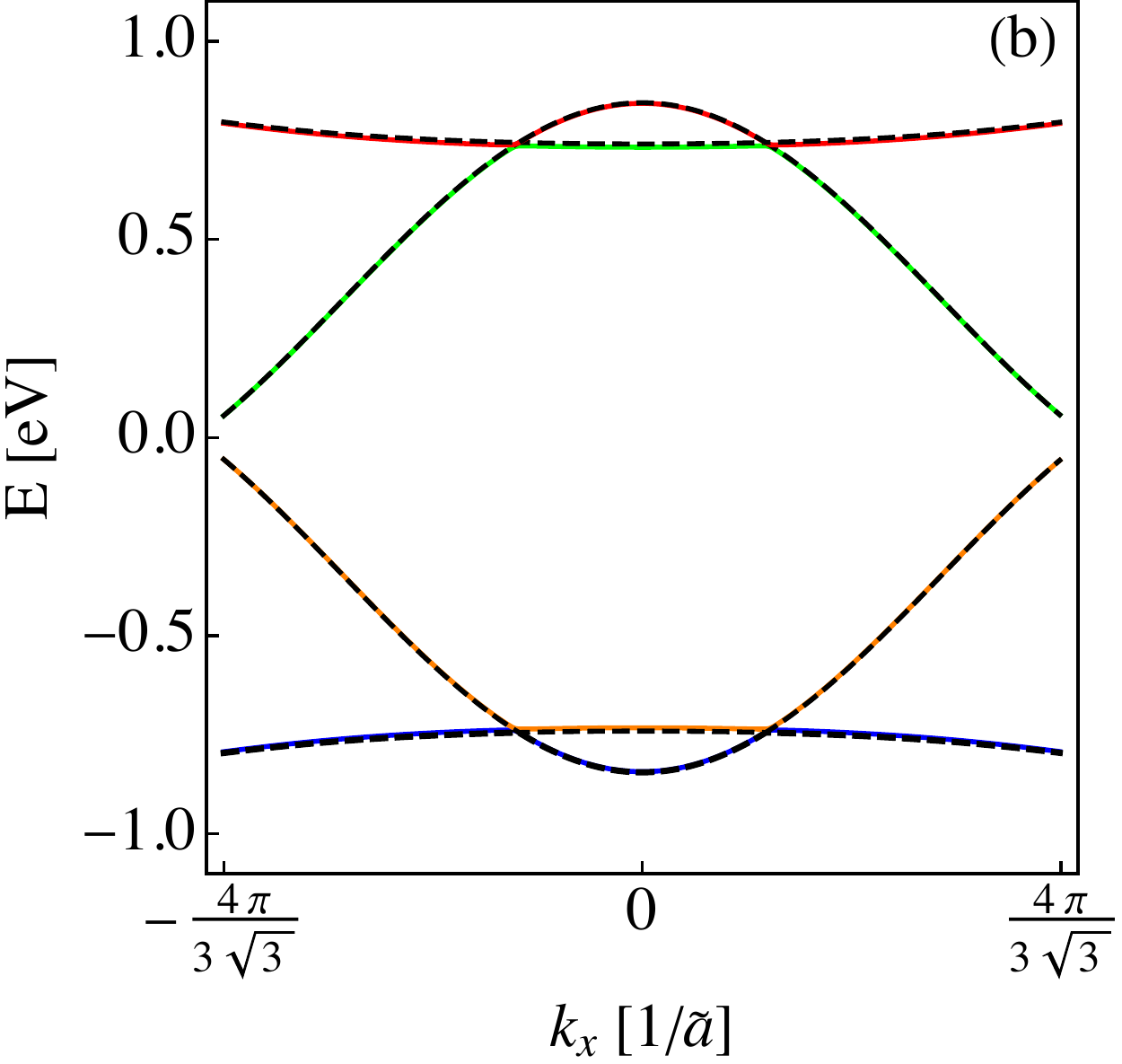}  %
\includegraphics[width=4cm]{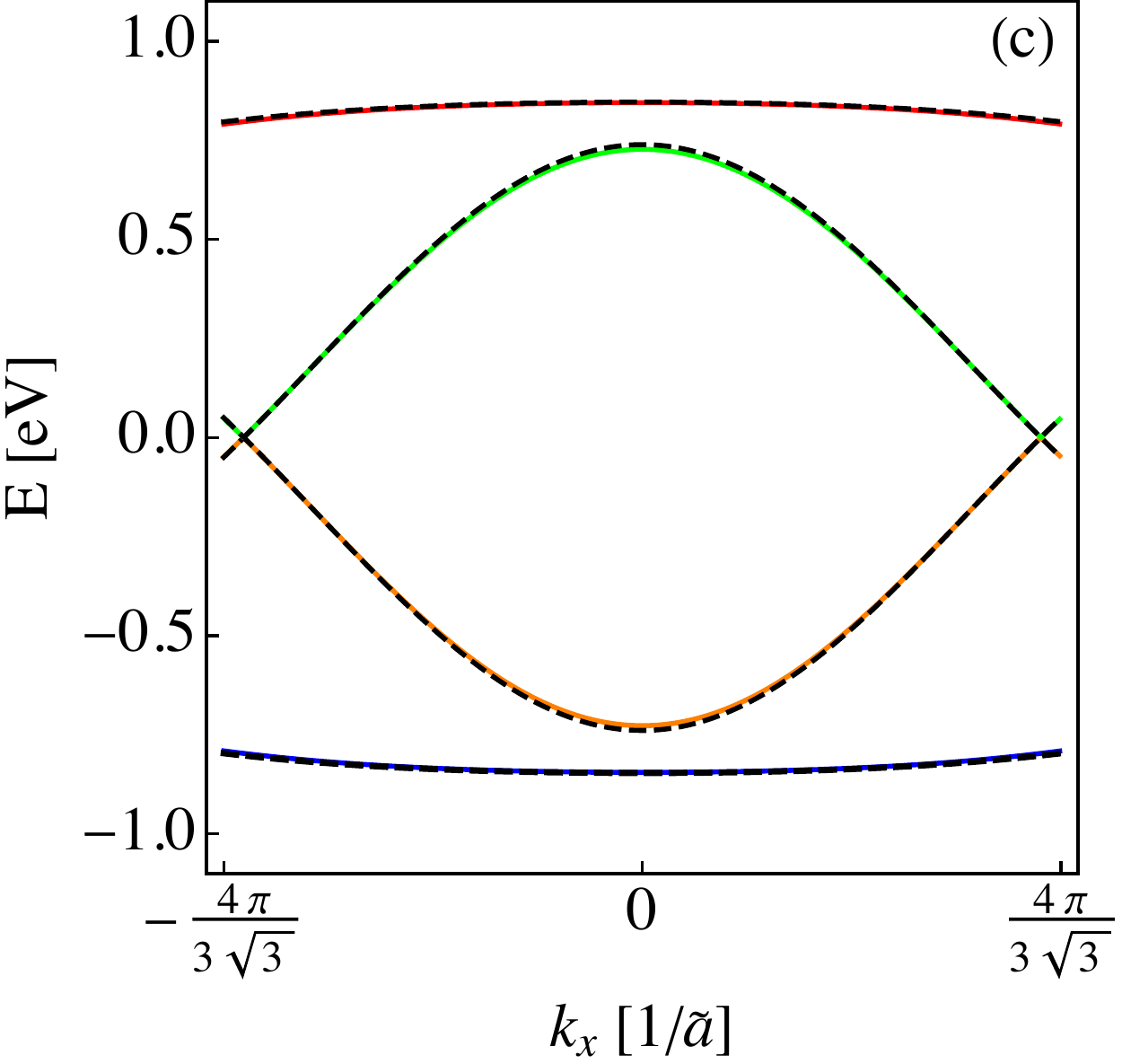}  %
\includegraphics[width=4cm]{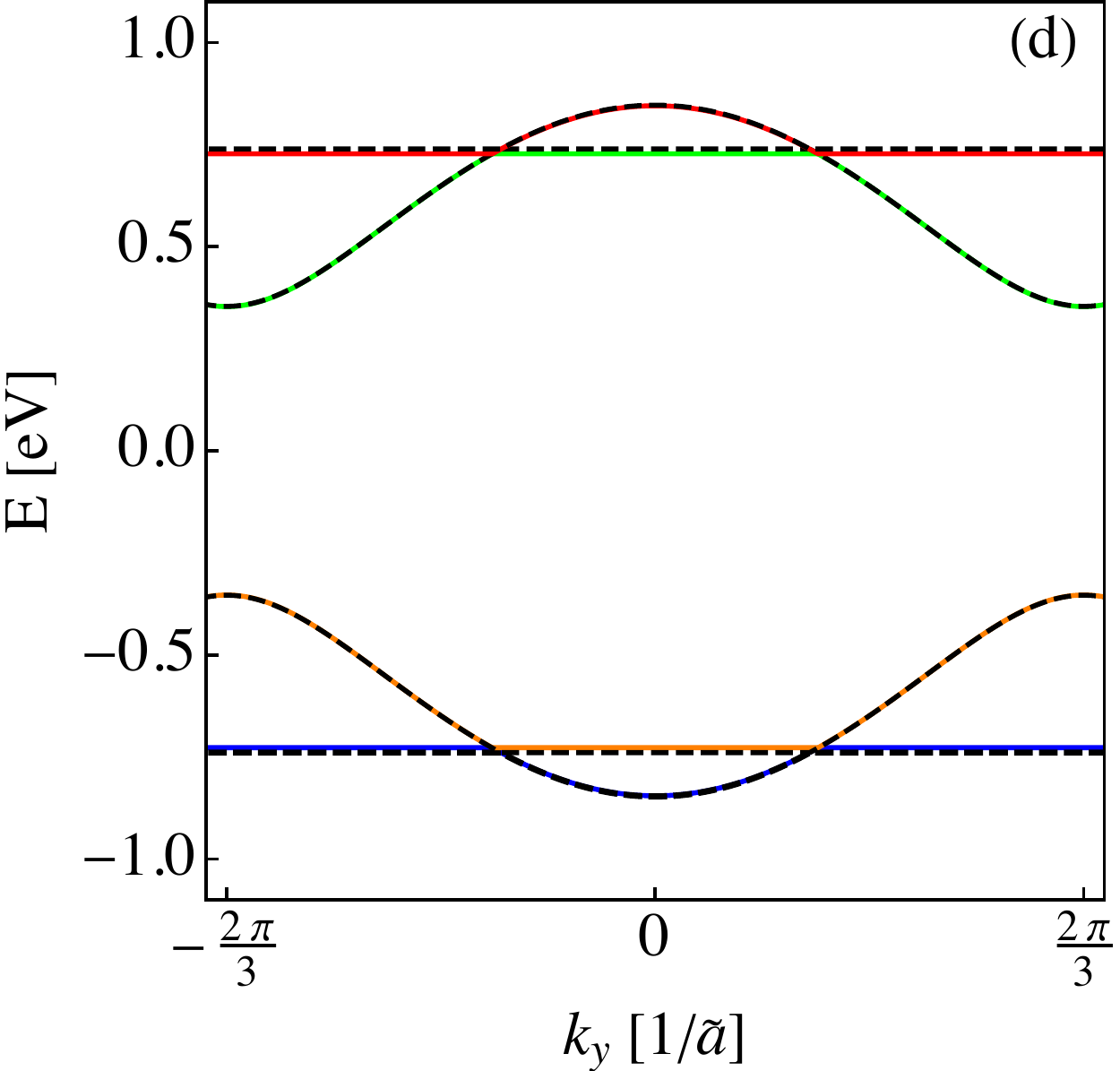}  %
\end{center}
\caption{(Color online) The Floquet-Bloch band structure of the (111) bilayer LaNiO$_3$ embedded in a normally incident linearly polarized light 
$\bfA(t) = A_0\cos(\Omega t)(\cos\theta, \sin\theta)$ with 
$\tilde{A}_0\tlda =1$ and $\hbar\Omega/t_\sigma=10$ (frequency $\nu=\Omega/2\pi=1.45079\times 10^3$THz and intensity $I=4.87887\times10^{8}$mW/$\mu$m$^2$). This is a high-energy off-resonant drive.
Only the dominant nearest neighbor hopping terms are kept in the tight-binding Hamiltonian Eq.\eqref{eq:htbk-t} 
with $t_\sigma=0.6$eV, $t'=0.0$eV.  The black dashed lines denote Floquet-Bloch band structure in the theoretical infinite frequency limit, i.e. the time-average Hamiltonian.
(a) The band structure plotted along the $k_x (k_y=0)$ direction in equilibrium (absence of laser), given by Eq.\eqref{eq:htbk}; 
(b) $\theta=\pi/2$, bands along. $k_x (k_y=0)$ direction.
(c)  $\theta=0$, bands along. $k_x (k_y=0)$ direction.
(d) $\theta=0$, bands along. $k_y (k_x=0)$ direction.
}
\label{fig:floquet-band-nn-lin}
\end{figure}

\begin{figure}[t]
\begin{center}
\includegraphics[width=4cm]{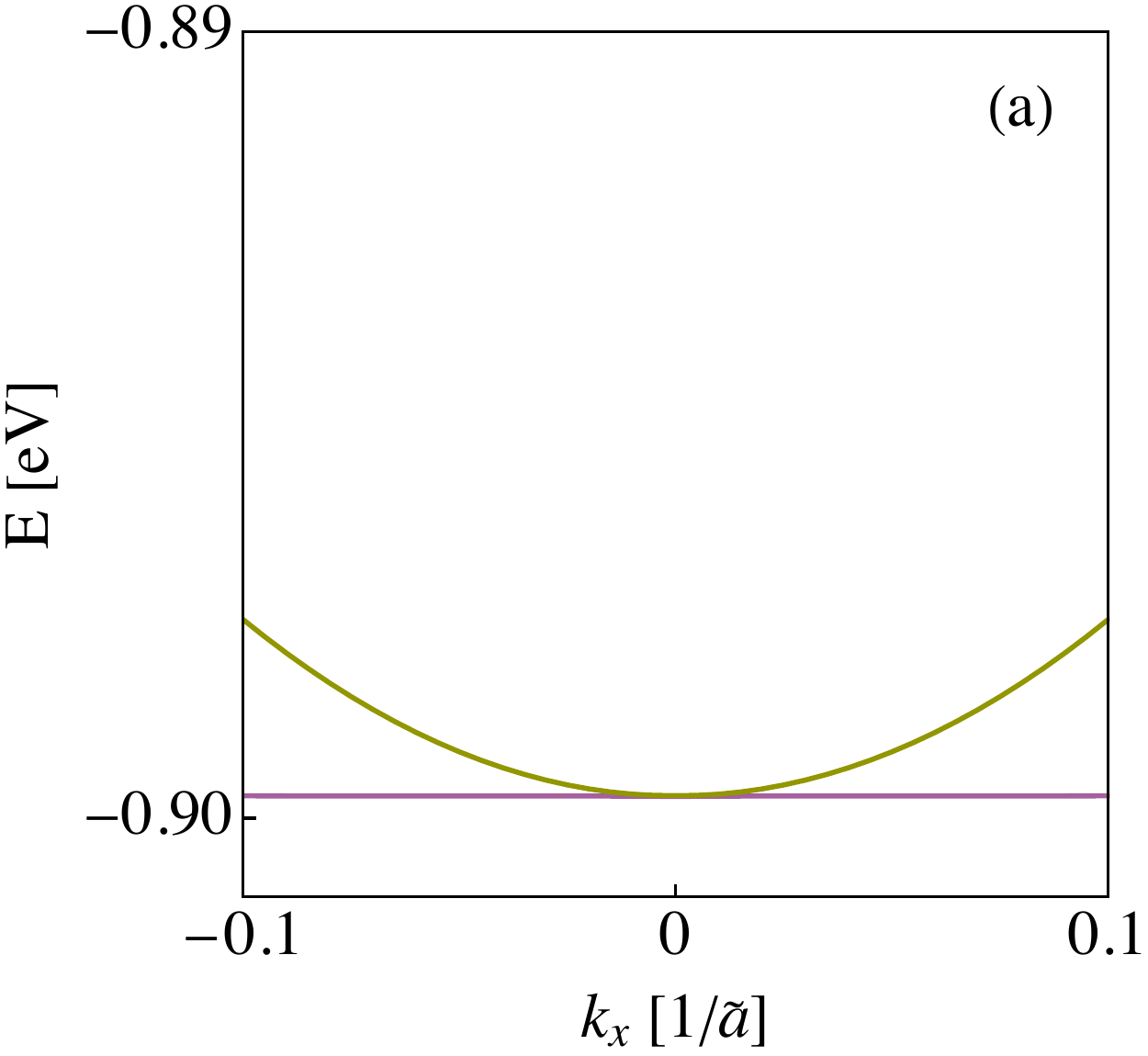}  %
\includegraphics[width=4cm]{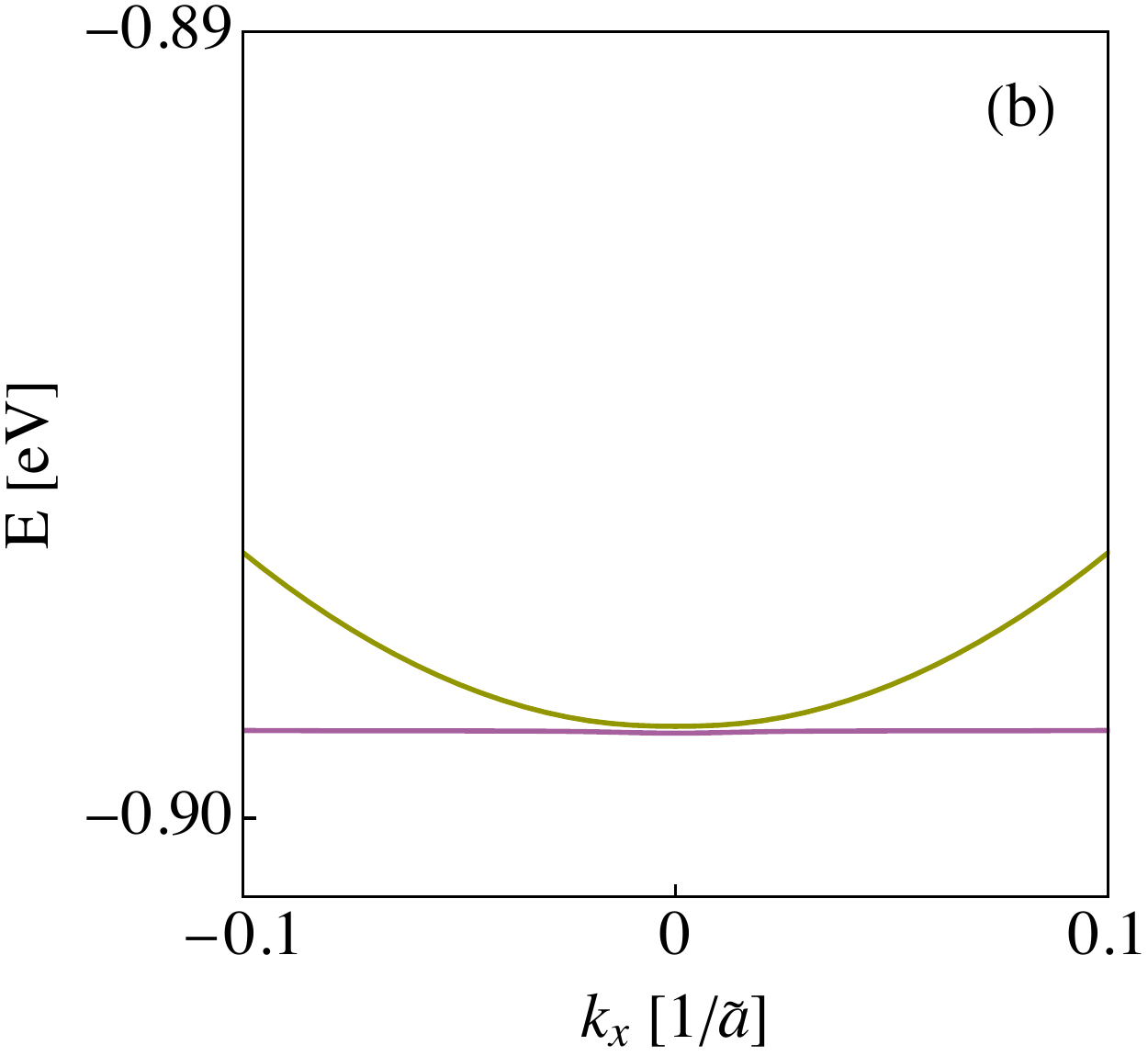}  %
\includegraphics[width=4cm]{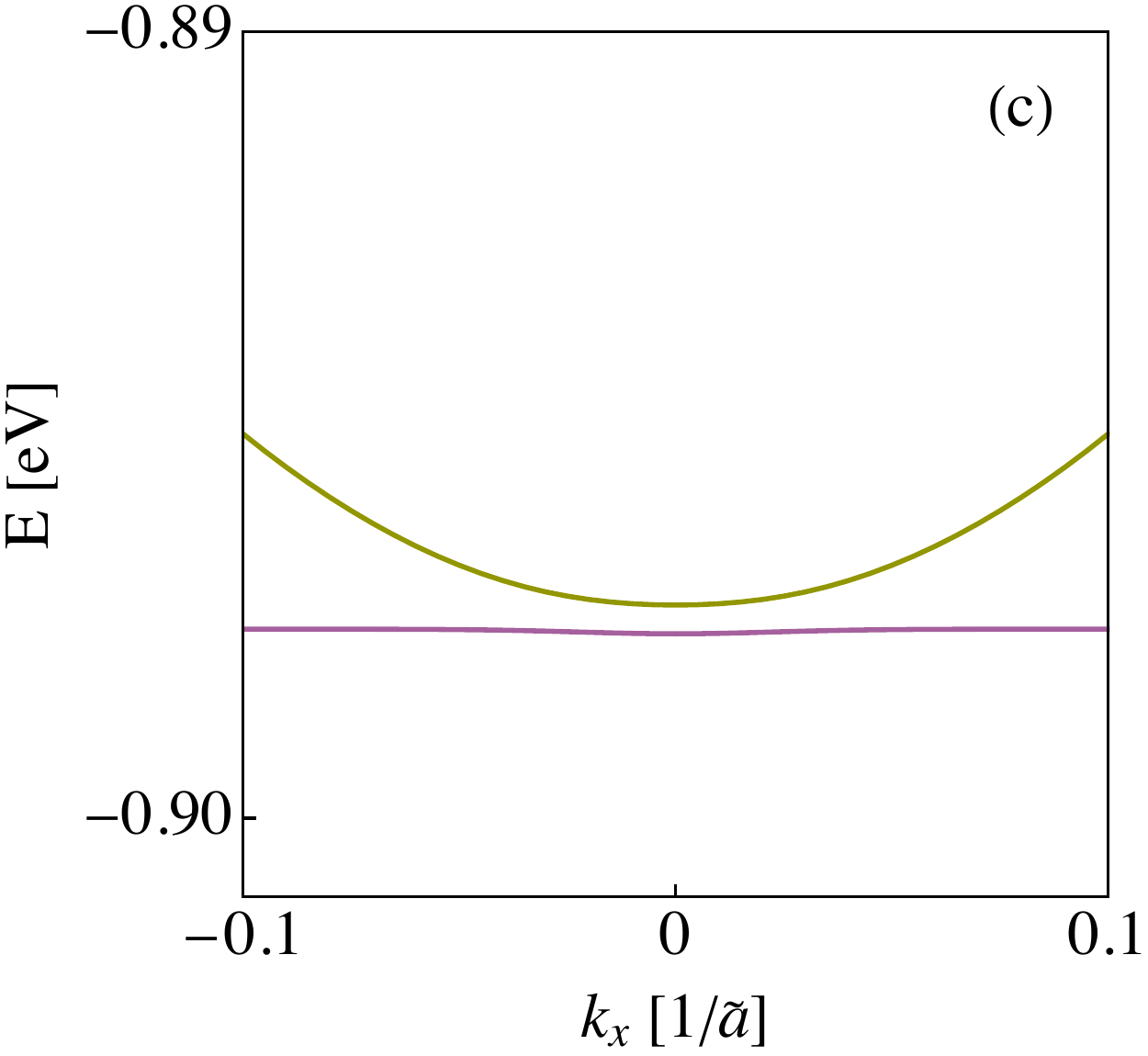}  %
\includegraphics[width=4cm]{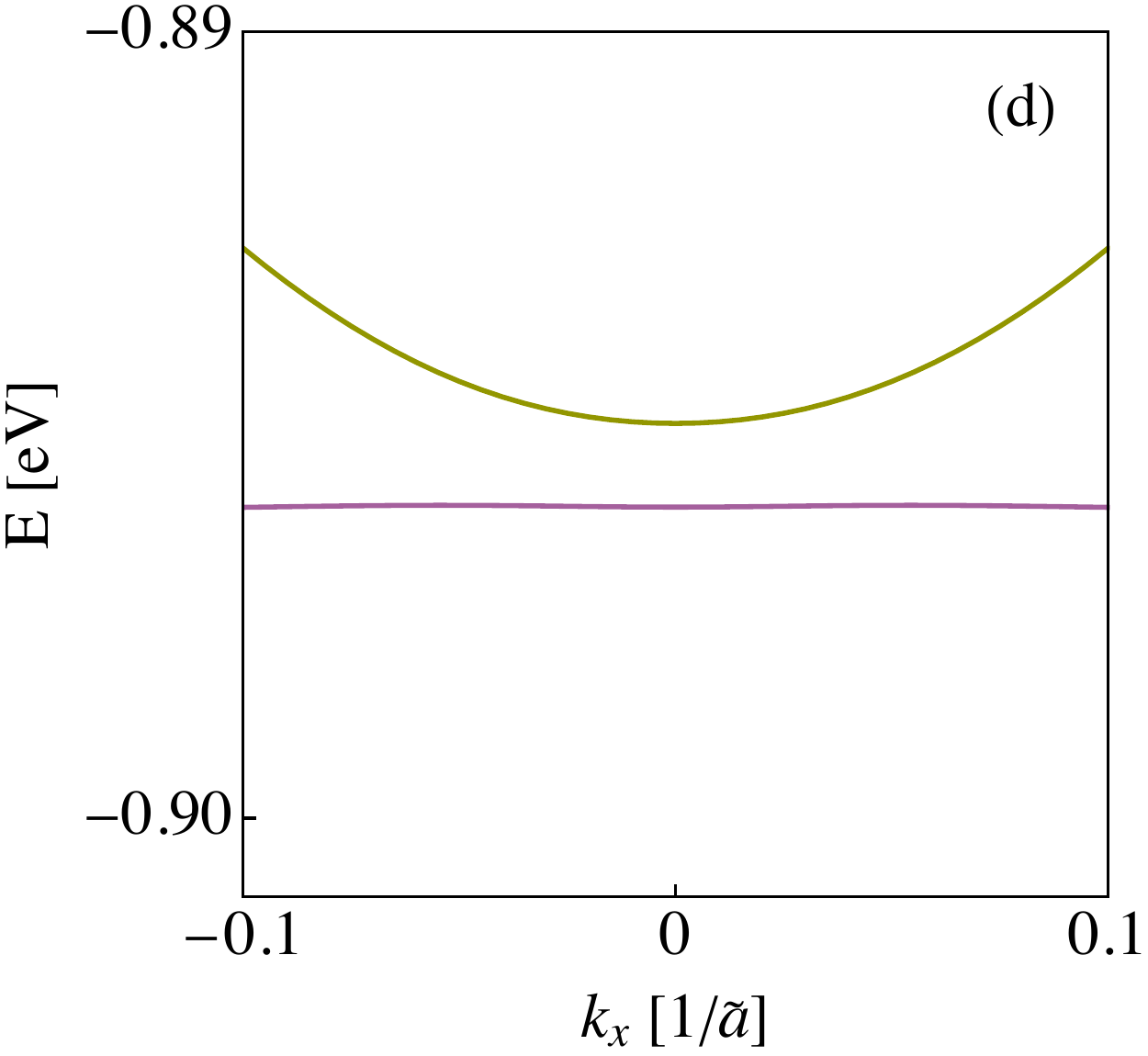}  %
\end{center}
\caption{(Color online) The Floquet-Bloch band structure of the (111) bilayer LaNiO$_3$ embedded in a normally incident circularly polarized light
$\bfA(t) = A_0(\cos\Omega t, -\sin\Omega t)$ with $\hbar\Omega=20.6783 meV$ (frequency $\nu=\Omega/2\pi=5$THz). This is a low-energy resonant drive.
Only the dominant nearest neighbor hopping terms are kept in the tight-binding Hamiltonian Eq.\eqref{eq:htbk-t} 
with $t_\sigma=0.6$eV, $t'=0.0$eV. 
(a) Bands along the $k_x (k_y=0)$ direction with $\tilde{A}_0 \tlda=0.05$, $I=14.57475$ mW/$\mu$m$^2$,
(b) Bands along the $k_x (k_y=0)$ direction with $\tilde{A}_0 \tlda=0.10$, $I=58.299$ mW/$\mu$m$^2$,
(c) Bands along the $k_x (k_y=0)$ direction with $\tilde{A}_0 \tlda=0.15$, $I=131.17275$ mW/$\mu$m$^2$, 
(d) Bands along the $k_x (k_y=0)$ direction with $\tilde{A}_0 \tlda=0.20$, $I=233.196$ mW/$\mu$m$^2$. 
}
\label{fig:floquet-band-nn-cir-THz}
\end{figure}

% figures with only nn hopping terms
\begin{figure}[t]
\begin{center}
\includegraphics[width=4cm]{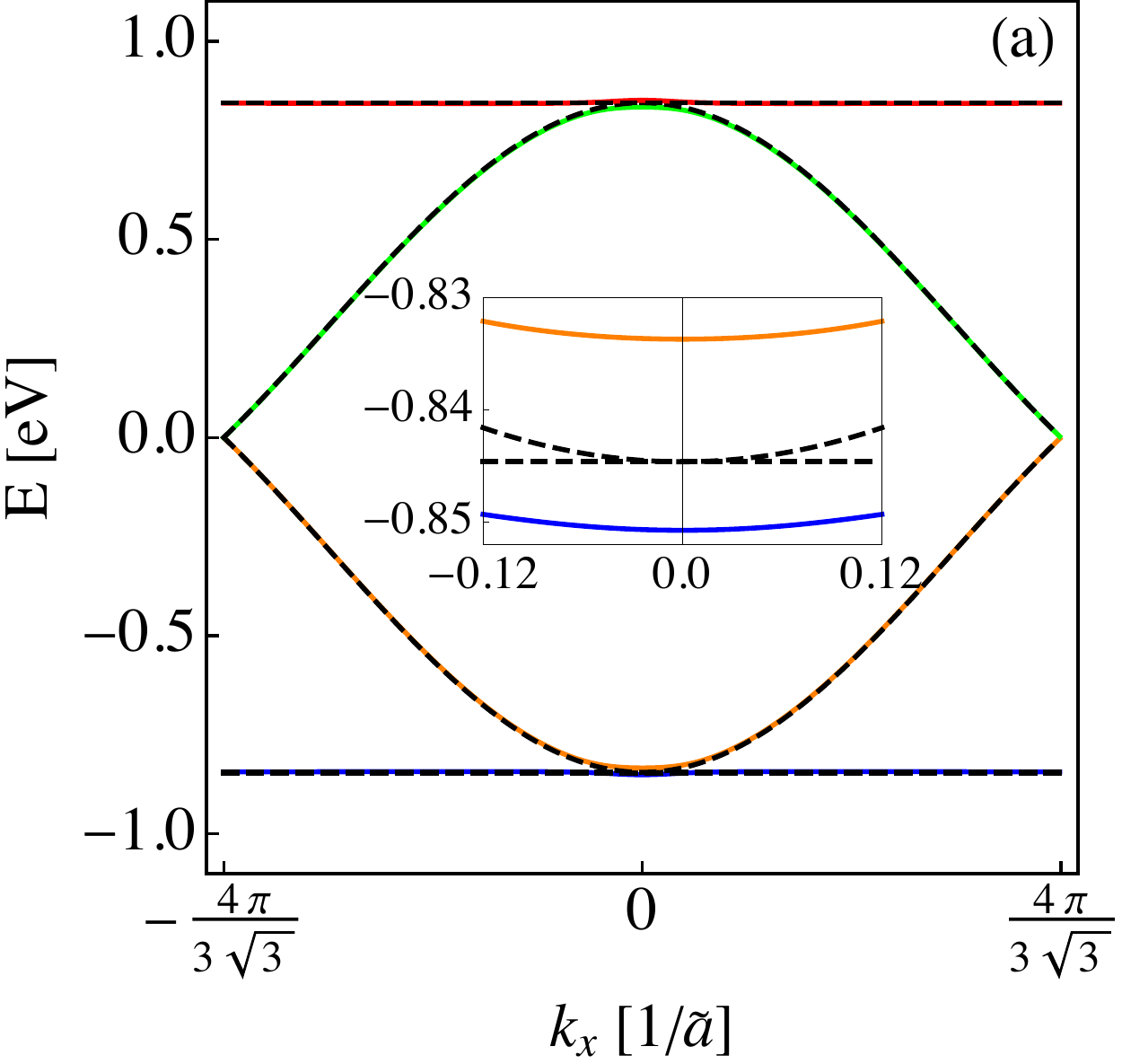}  %
\includegraphics[width=4cm]{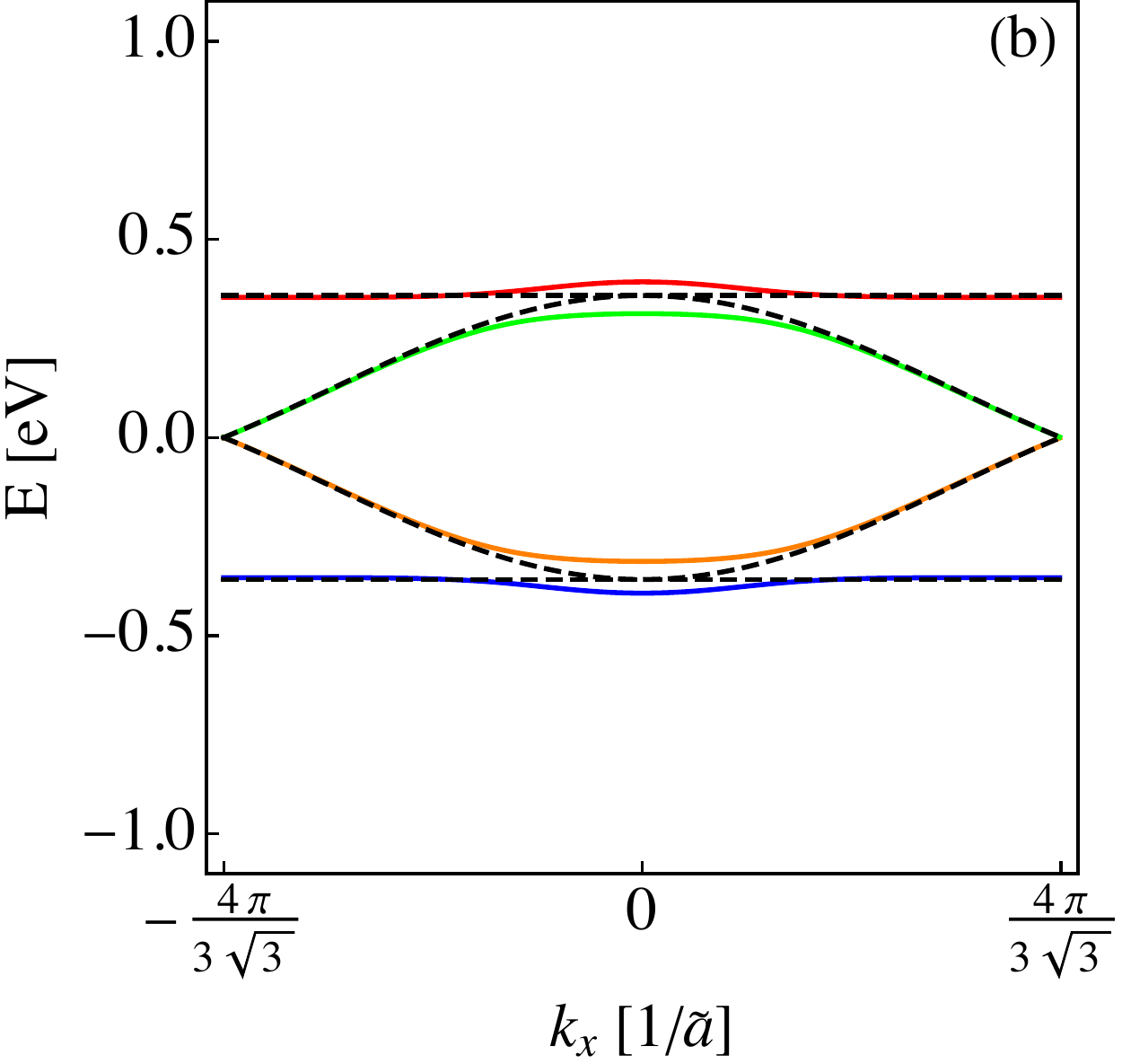}  %
\includegraphics[width=4cm]{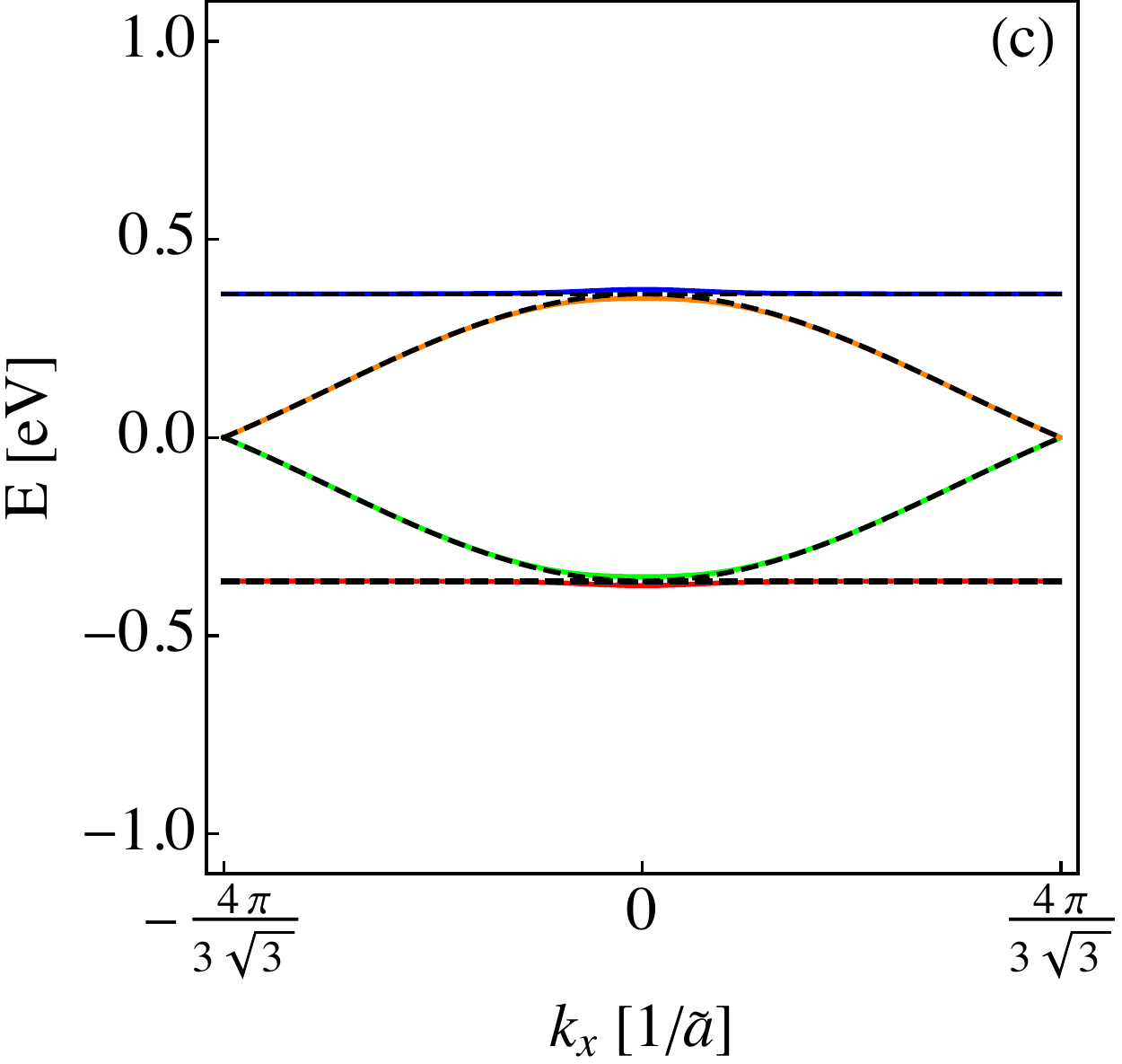}  %
\includegraphics[width=4cm]{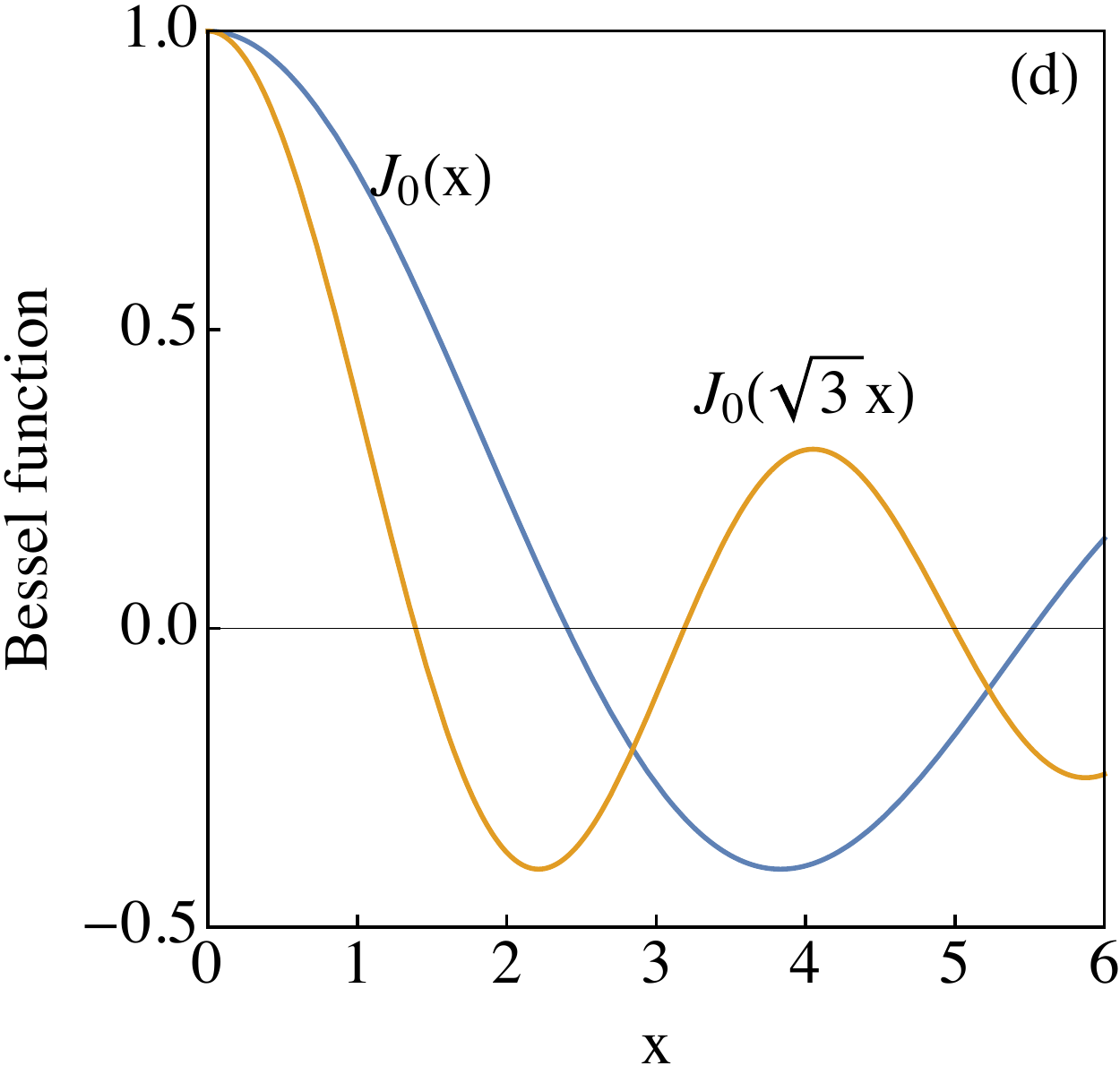}  %
\end{center}
\caption{(Color online) The Floquet-Bloch band structure of the (111) bilayer LaNiO$_3$ embedded in a normally incident circularly polarized light
$\bfA(t) = A_0(\cos\Omega t, -\sin\Omega t)$ with $\hbar\Omega/t_\sigma=10$ (frequency $\nu=\Omega/2\pi=1.45079\times 10^3$THz).  This is a high-energy off-resonant drive. Only the dominant nearest neighbor hopping terms are kept in the tight-binding Hamiltonian Eq.\eqref{eq:htbk-t} with $t_\sigma=0.6$eV, $t'=0.0$eV.  The black dashed lines denote Floquet-Bloch band structure in the theoretical infinite frequency limit.
(a) Bands along the $k_x (k_y=0)$ direction with $\tilde{A}_0 \tlda=0.5$, $I=1.21972\times10^{8}$mW/$\mu$m$^2$,
the inset shows a zoomed view around the quadratic band touching at 1/4 filling,
(b) Bands along the $k_x (k_y=0)$ direction with $\tilde{A}_0 \tlda=1.7$, $I=1.40999\times10^{9}$mW/$\mu$m$^2$,
(c) Bands along the $k_x (k_y=0)$ direction with  $\tilde{A}_0 \tlda=3.8$, $I=7.04509\times10^{9}$mW/$\mu$m$^2$), 
(d) The zero-$th$ order Bessel function of the first kind is plotted as $\mathcal{J}_0(x)$ and $\mathcal{J}_0(\sqrt{3}x)$ responsible for independent renormalization of first and second neighbor hopping.
}
\label{fig:floquet-band-nn-cir}
\end{figure}

% figures with both nn and nnn hopping terms
\begin{figure}[h]
\begin{center}
\includegraphics[width=4cm]{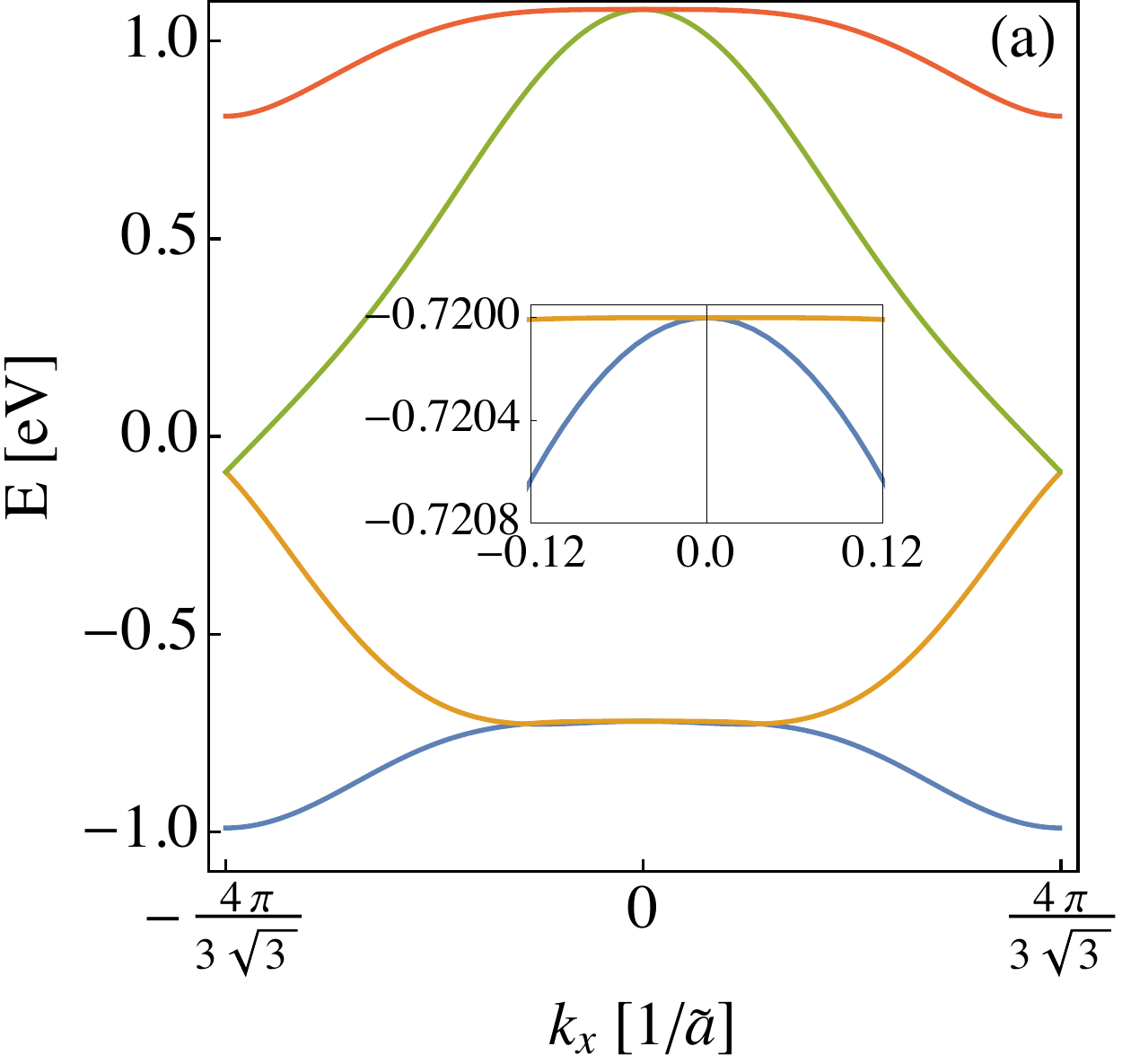}  %
\includegraphics[width=4cm]{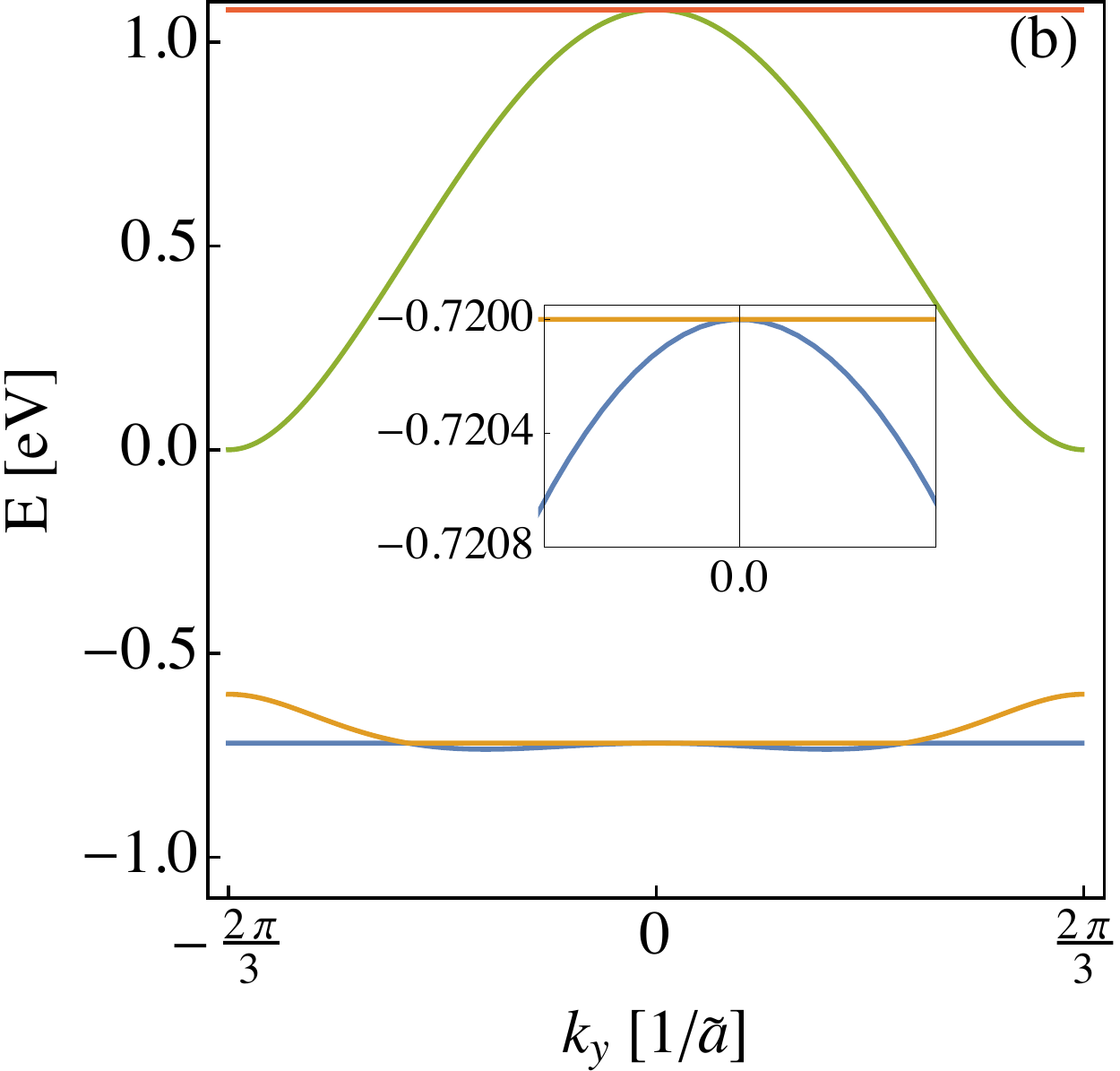}  %
\includegraphics[width=4cm]{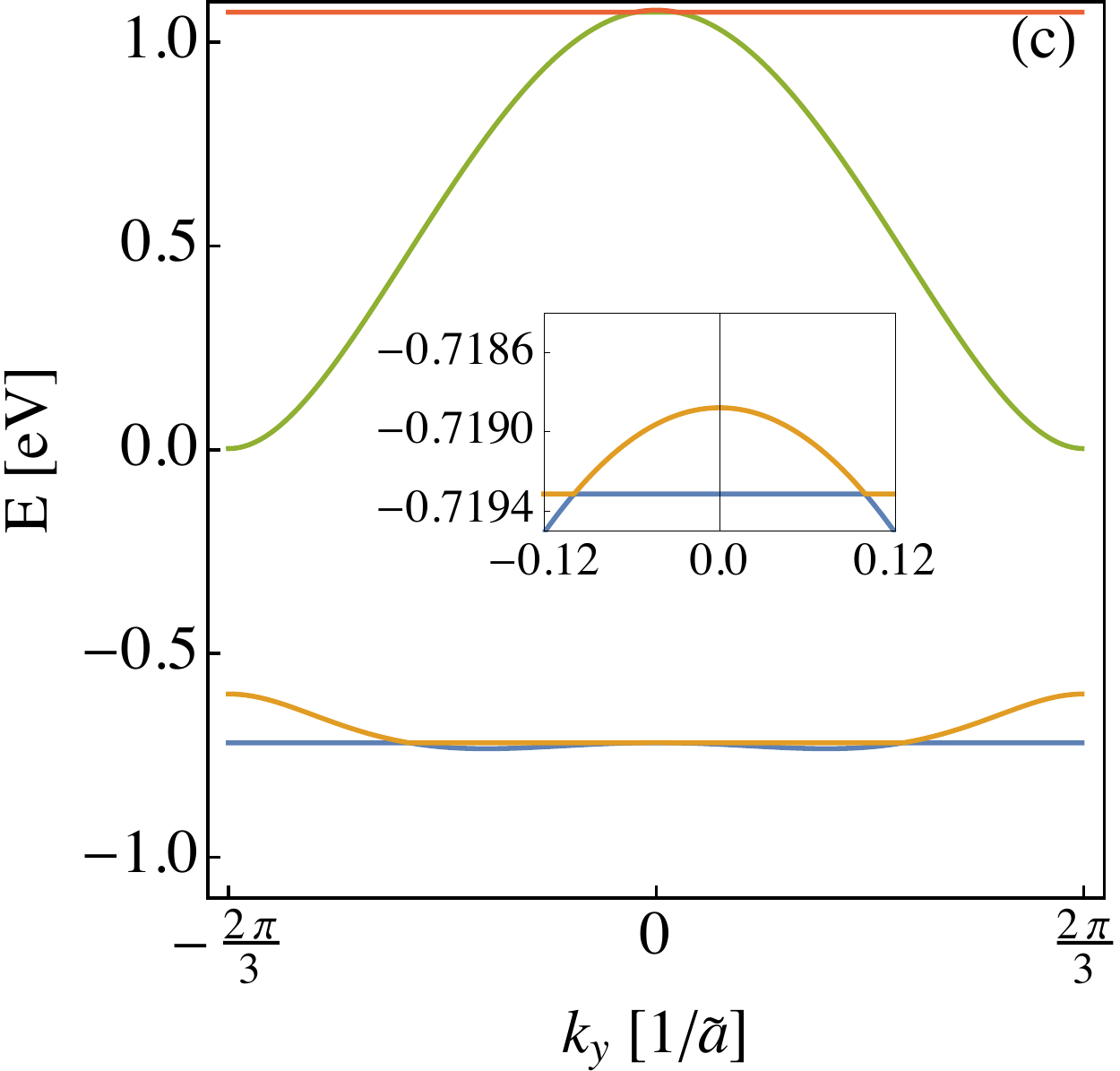}  %
\includegraphics[width=4cm]{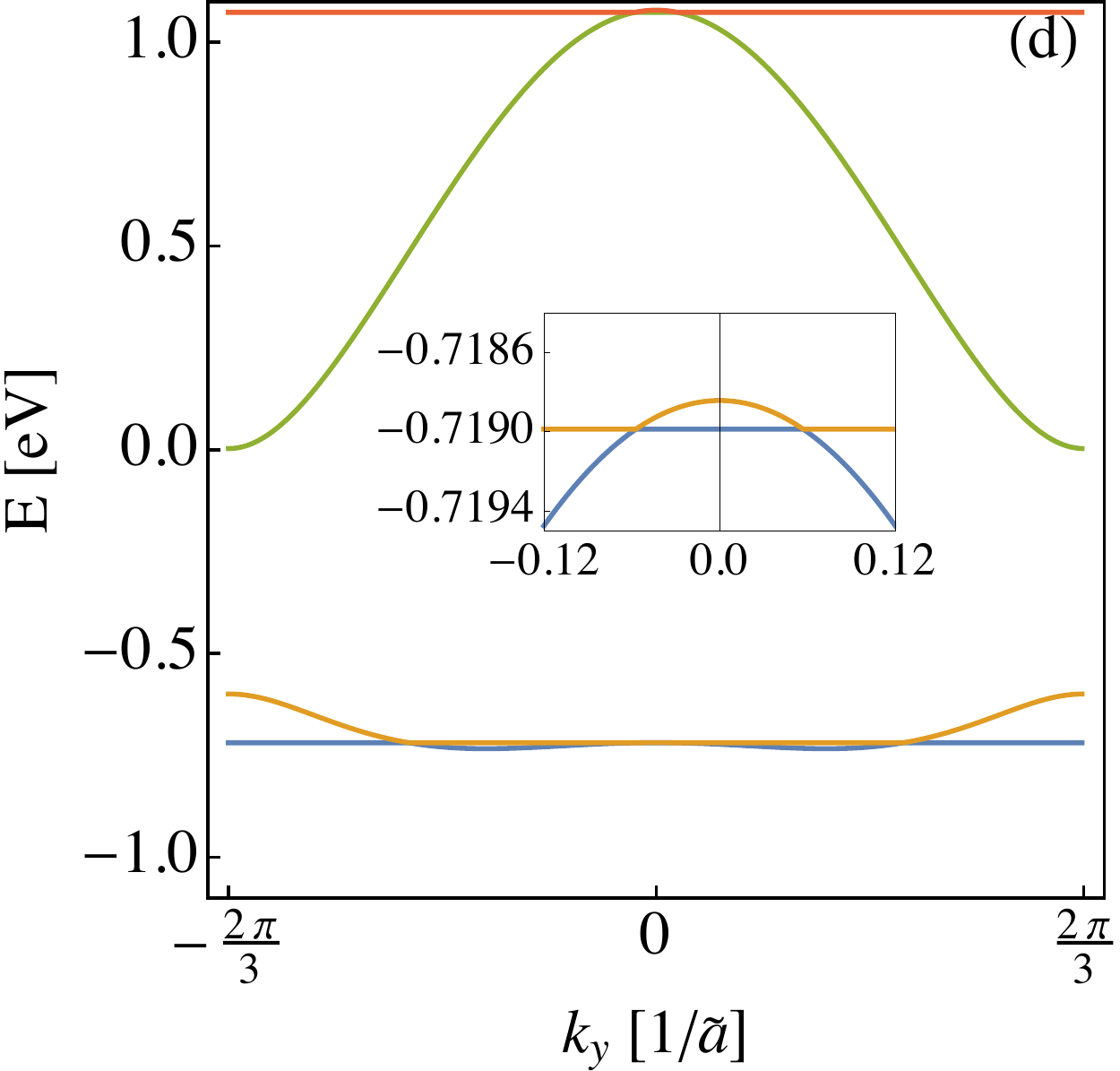}  %
\includegraphics[width=4cm]{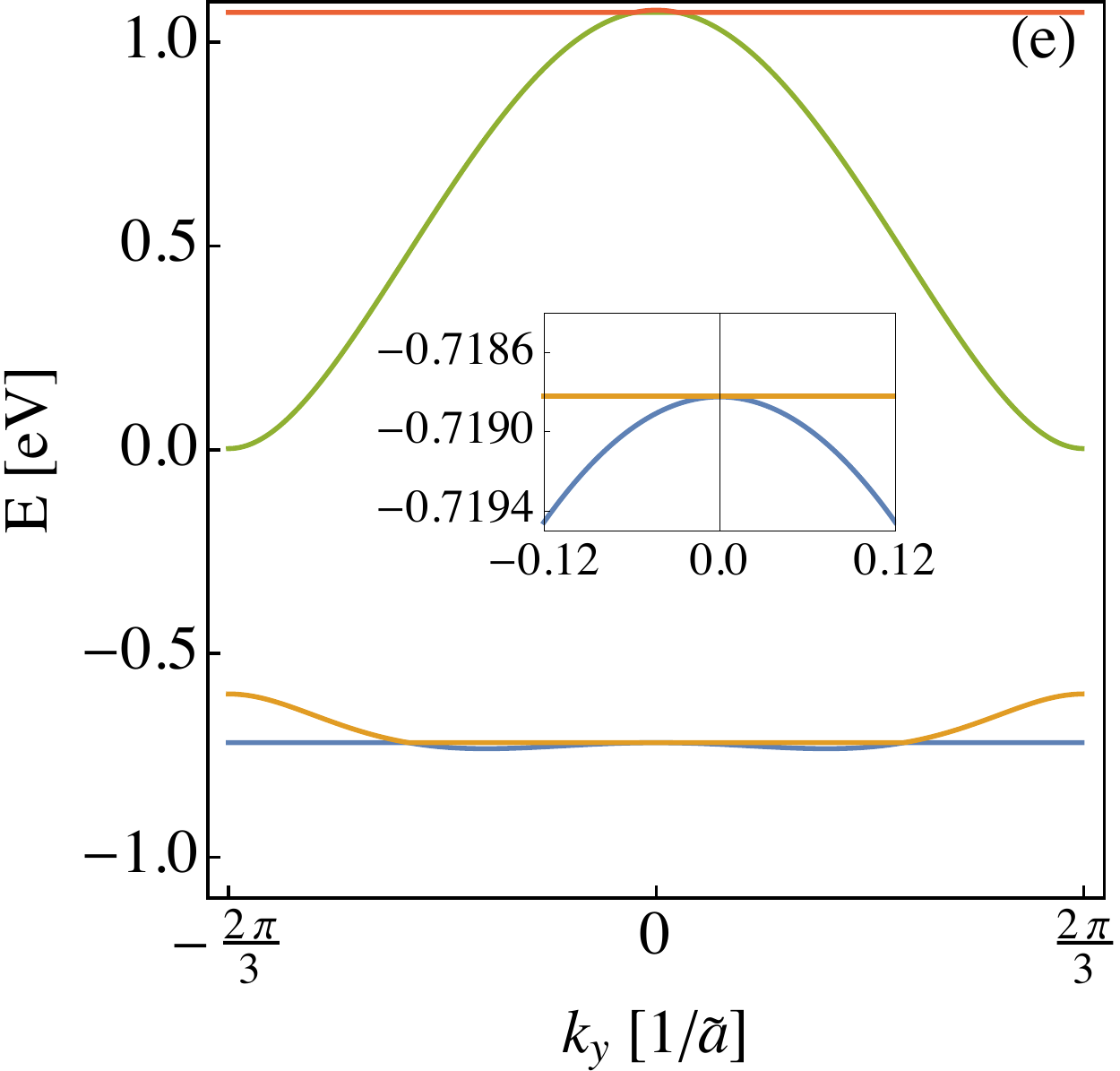}  %
\includegraphics[width=4cm]{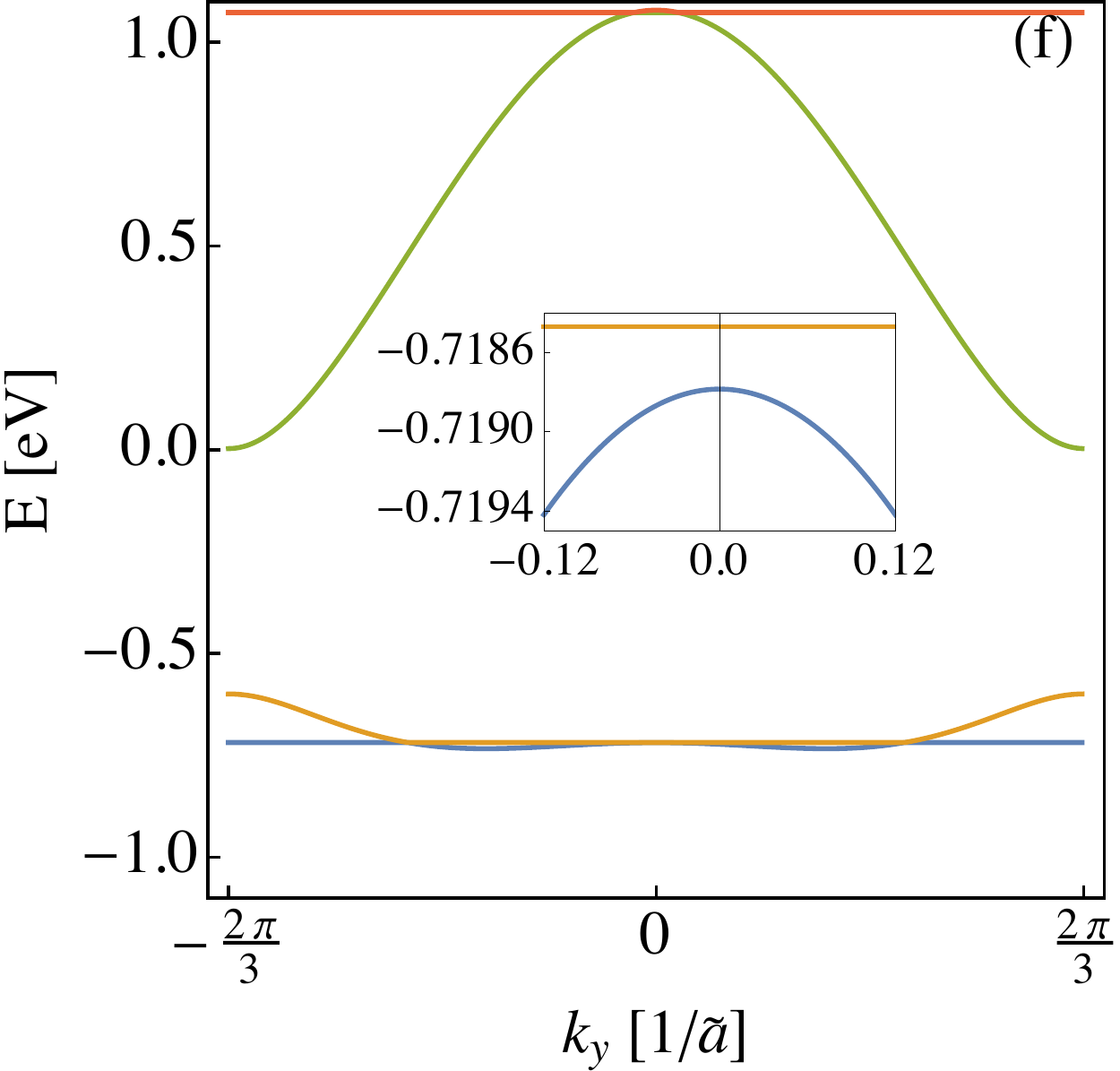}  %
\end{center}
\caption{(Color online) The Floquet-Bloch band structure of the (111) bilayer LaNiO$_3$ embedded in a normally incident linearly polarized light
$\bfA(t) = A_0\cos(\Omega t)(\cos\theta, \sin\theta)$ with 
$\tilde{A}_0\tlda=\sqrt{2}/10$. The series shows the approach to a resonant drive starting from a  high-energy off-resonant drive. The electric field is fixed to point along the $x$ direction by $\theta=0$. Both nearest neighbor and next nearest neighbor hopping terms are kept in the tight-binding Hamiltonian Eq.\eqref{eq:htbk-t} with $t_\sigma=0.6$eV, $t'=0.1t_\sigma$. 
The band structure plotted along: (a) $k_x (k_y=0)$; (b) $k_y (k_x=0)$ direction in equilibrium (absence of laser), given by Eq.\eqref{eq:htbk}, 
(c) $k_y (k_x=0)$ direction, $\hbar\Omega/t_{\sigma} = 100$, $\nu=14.5079\times 10^3$THz, $I=9.75774\times10^{8}$mW/$\mu$m$^2$; 
(d) $k_y (k_x=0)$ direction, $\hbar\Omega/t_{\sigma} = 10.0$, $\nu=1.45079\times 10^3$THz, $I=9.75774\times10^{6}$mW/$\mu$m$^2$; 
(e) $k_y (k_x=0)$ direction, $\hbar\Omega/t_{\sigma} = 8.33$, $\nu=1.20899\times 10^3$THz, $I=6.77621\times10^{6}$mW/$\mu$m$^2$;
(f) $k_y (k_x=0)$ direction, $\hbar\Omega/t_{\sigma} = 6.67$,  $\nu=0.96719\times 10^3$THz, $I=4.33677\times10^{6}$mW/$\mu$m$^2$. All insets show a zoomed view around the quadratic touching at $1/4$ filling which shows the most signifiant renormalization.
}
\label{fig:floquet-band-nnn-lin}
\end{figure}

% figures with both nn and nnn hopping terms
\begin{figure}[th]
\begin{center}
\includegraphics[width=4cm]{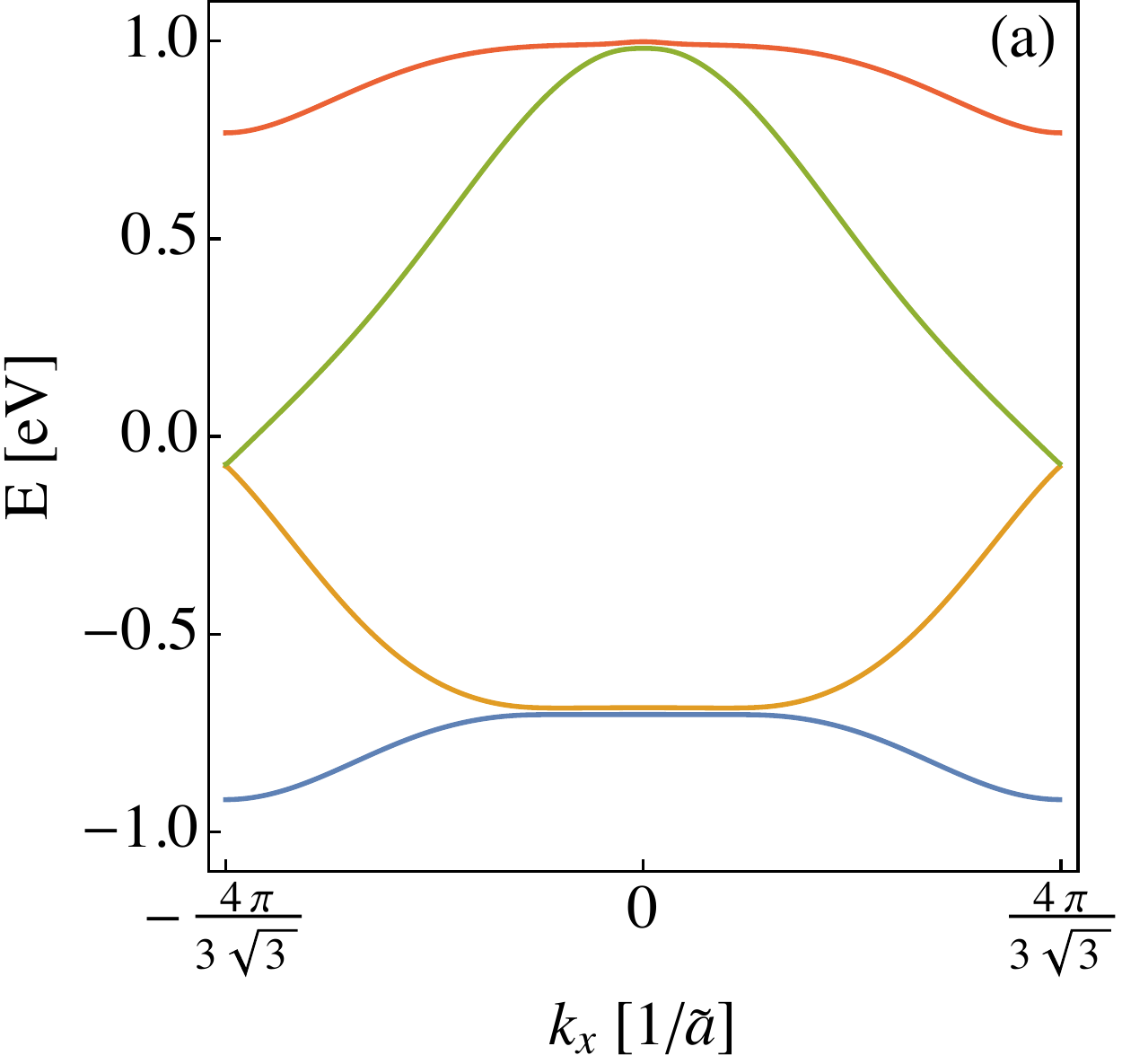}  %
\includegraphics[width=4cm]{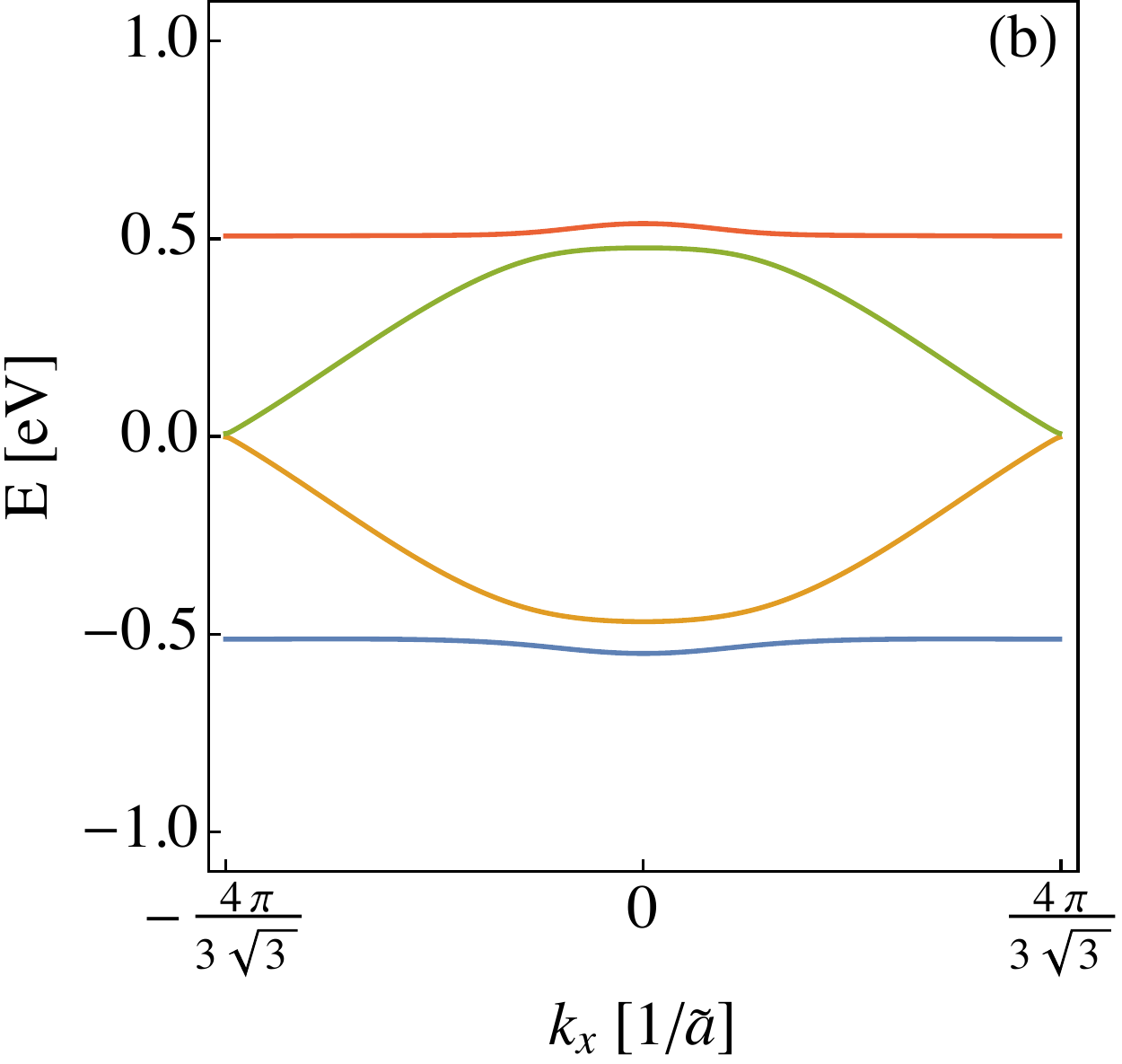}  %
\includegraphics[width=4cm]{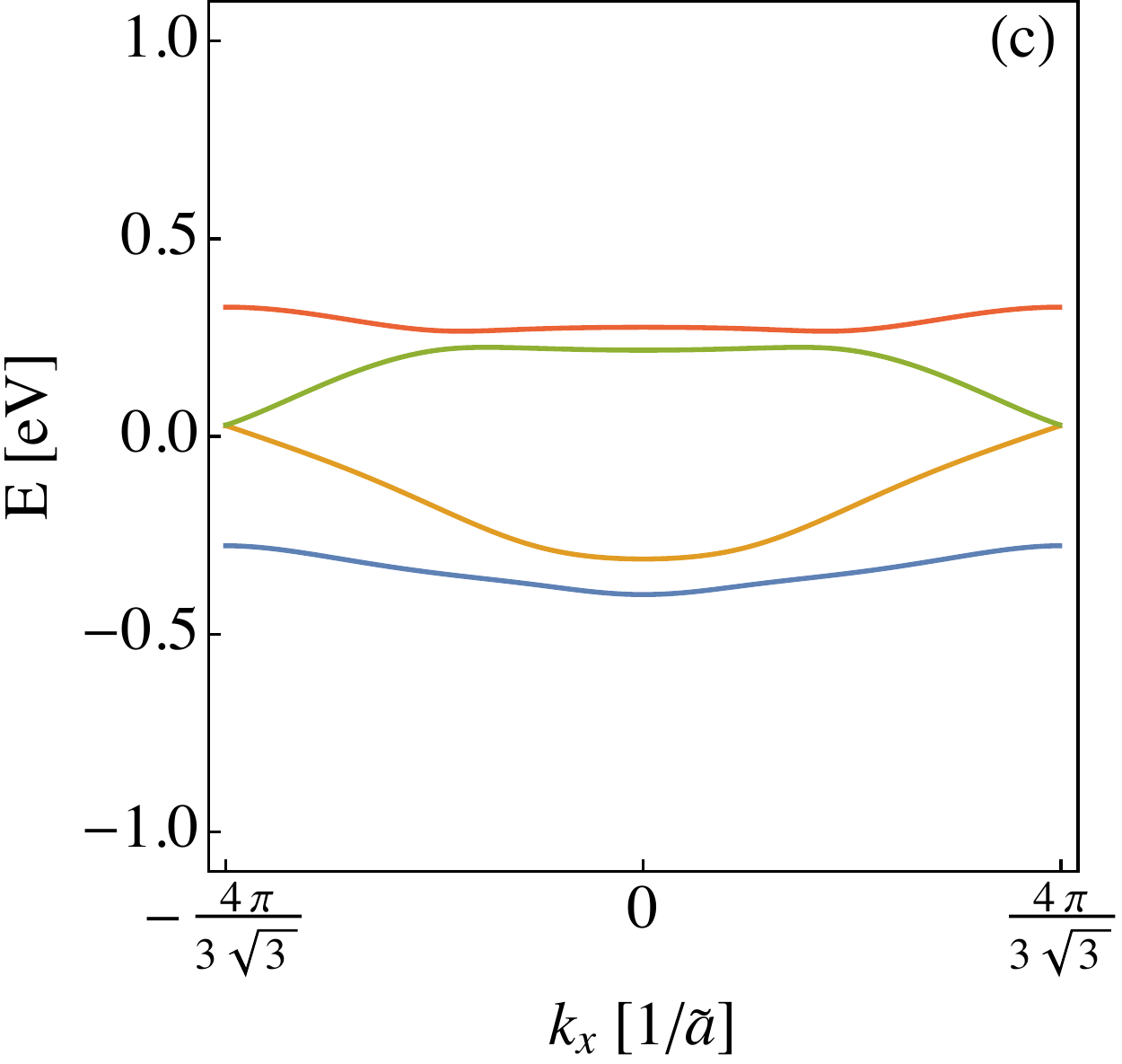}  %
\includegraphics[width=4cm]{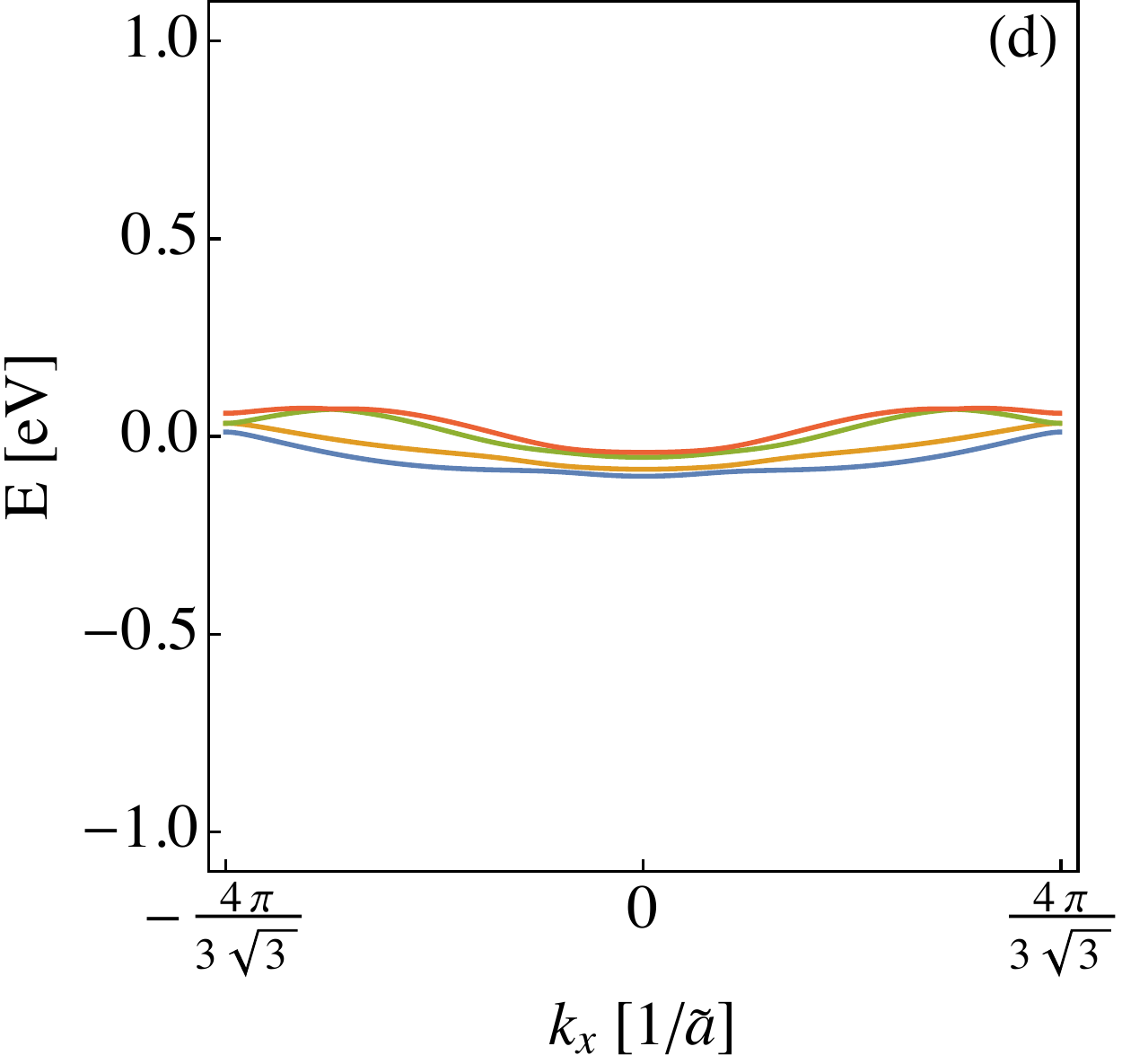}  %
\includegraphics[width=4cm]{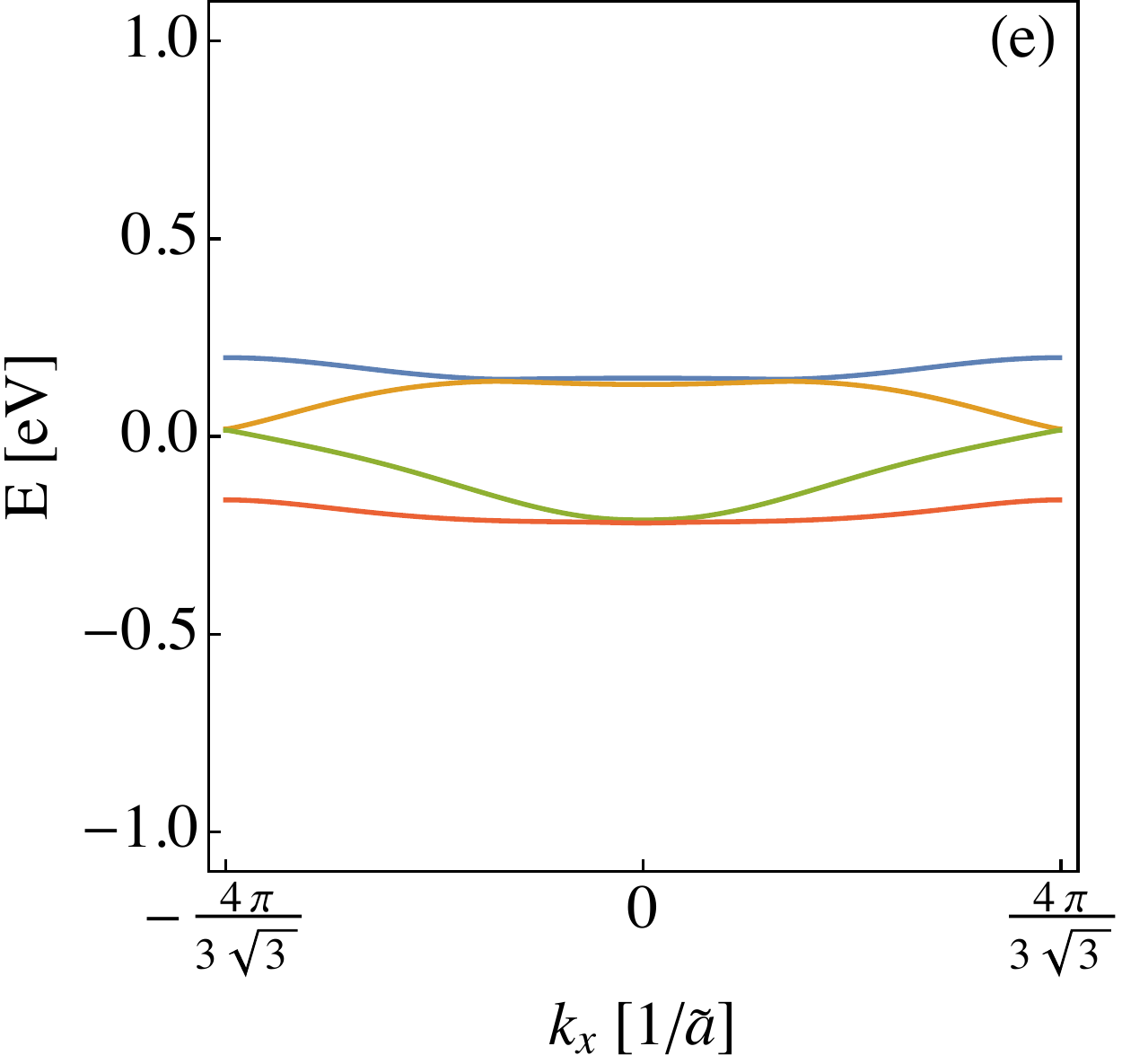}  %
\includegraphics[width=4cm]{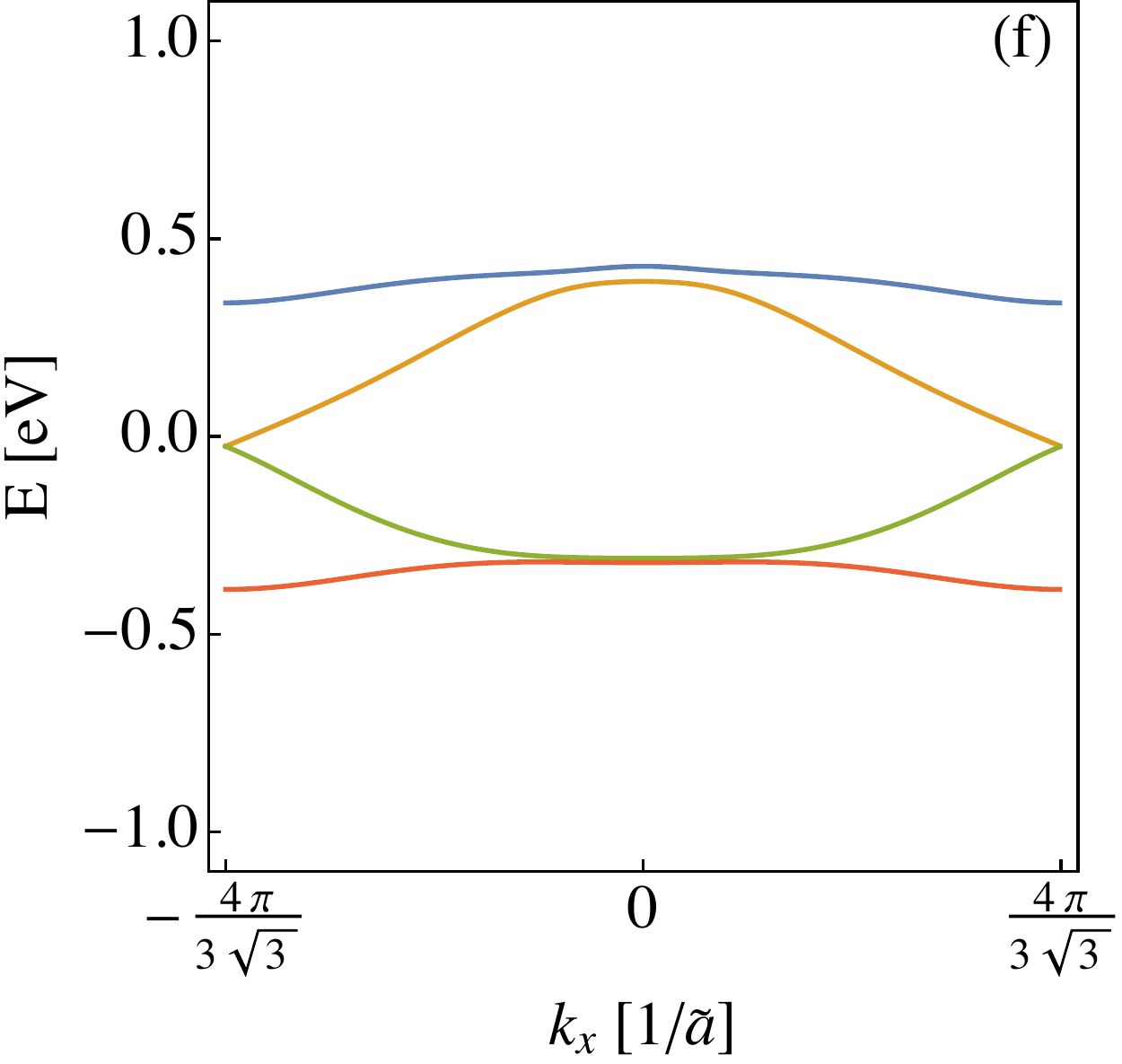}  %
\end{center}
\caption{(Color online) The Floquet-Bloch band structure of the (111) bilayer LaNiO$_3$ embedded in a normally incident circularly polarized light
$\bfA(t) = A_0(\cos\Omega t, -\sin\Omega t)$ with $\hbar\Omega/t_\sigma=10$ (frequency $\nu=\Omega/2\pi=1.45079\times 10^3$THz). Series shows off-resonant drive with increasing intensity.  Both nearest neighbor and next nearest neighbor hopping terms are kept in the tight-binding Hamiltonian Eq.\eqref{eq:htbk-t} 
with $t_\sigma=0.6$eV, $t'=0.1t_\sigma$.
(a) $k_x (k_y=0)$ direction, circularly polarized light $\tilde{A}_0 \tlda=0.50$, $I=1.21972\times10^{8}$mW/$\mu$m$^2$;
(b) $k_x (k_y=0)$ direction, circularly polarized light $\tilde{A}_0 \tlda=1.39$, $I=9.42646\times10^{8}$mW/$\mu$m$^2$;
(c) $k_x (k_y=0)$ direction, circularly polarized light $\tilde{A}_0 \tlda=1.80$, $I=1.58075\times10^{9}$mW/$\mu$m$^2$; 
(d) $k_x (k_y=0)$ direction, circularly polarized light $\tilde{A}_0 \tlda=2.40$, $I=2.81023\times10^{9}$mW/$\mu$m$^2$; 
(e) $k_x (k_y=0)$ direction, circularly polarized light $\tilde{A}_0 \tlda=2.83$, $I=3.90744\times10^{9}$mW/$\mu$m$^2$; 
(f)  $k_x (k_y=0)$ direction, circularly polarized light $\tilde{A}_0 \tlda=3.80$, $I=7.04509\times10^{9}$mW/$\mu$m$^2$.
}
\label{fig:floquet-band-nnn-cir}
\end{figure}

\subsection{Linearly Polarized Light}
For linearly polarized light with vector potential $\bfA(t)=A_0 \cos(\Omega t)[\cos\theta, \sin\theta]$, the Floquet Hamiltonian matrix elements are
\begin{align}
     H^{AB}_{nm} &= t_x g_0(-k_2, \gamma_1) + t_y g_0(-k_3, \gamma_2) + t_z g_0(k_0, \gamma_3), \nonumber\\
     H^{BA}_{nm} &= t_x^* g_0(k_2,-\gamma_1) + t_y^* g_0(k_3, -\gamma_2) + t_z^* g_0(k_0,-\gamma_3), \nonumber\\
     H^{AA}_{nm} &= t_{xy} [g_1(-k_1, \gamma_3, \tilde{A}_0) + g_1(+k_1, -\gamma_3, -\tilde{A}_0)] \nonumber\\
                 &+ t_{zx} [g_1(-k_2, \gamma_2, \tilde{A}_0) + g_1(+k_2, -\gamma_2, -\tilde{A}_0)] \nonumber\\
                 &+ t_{yz} [g_1(-k_3, \gamma_1, \tilde{A}_0) + g_1(+k_3, -\gamma_1, -\tilde{A}_0)],
\end{align}
where 
$g_1(k_i, \gamma, x) = i^{m-n} e^{i \bfk \cdot \bfa_i} \mathcal{J}_{m-n}(\sqrt{3}x\tlda\cos\gamma)$ and
$g_0(k_i, \gamma) = i^{m-n} e^{i \bfk \cdot \bfa_i} \mathcal{J}_{m-n}(\tilde{A}_0\tlda\sin\gamma)$ with
$\gamma_1 = \theta-2\pi/3, \gamma_2=\theta-\pi/3, \gamma_3=\theta$, and 
$g_0(k_0, \gamma) = \mathcal{J}_{m-n}(\tilde{A}_0\tlda\sin\gamma) i^{m-n}$.  Here $t_u^*$ is the complex conjugate of $t_u$, and 
$\mathcal{J}_m(x)$ is the order-$m$ Bessel function of the first kind, as before. 

The Floquet-Bloch band structure can be derived by diagonalizing the Floquet Hamiltonian in the truncated Floquet space (restricted values of $m$). We first consider the case where only the dominant nearest-neighbor hopping term $t_\sigma=0.6$eV is taken into account.  The equilibrium band structure is shown in Fig.\ref{fig:floquet-band-nn-lin}(a).  Focusing in on the response of the quadratic band touching point at the $\Gamma$ point, Fig.\ref{fig:floquet-band-nn-lin-THz} shows the band structure ($t'/t_\sigma=0.0$) for linearly polarized light 
with fixed amplitude $\tilde{A}_0\tlda =\sqrt{2}/10$ and frequency $\hbar\Omega=0.0206783$ eV. These parameters are experimentally realizable, and correspond to a ``resonant" case with driving energy well-below the band width.   

By setting the polarization direction to be along the $x$-direction, we plot the band structure along $k_x (k_y=0)$ and $k_y (k_x=0)$ in Fig.\ref{fig:floquet-band-nn-lin-THz} (a-b). The quadratic touching point will split into two Dirac points in $k_x$ direction and gapped along $k_y$ direction.
The band structure with different laser polarization direction (along $y$) are plotted in Fig.\ref{fig:floquet-band-nn-lin-THz} (c-d). The quadratic touching point is gapped along both $k_x$ and $k_y$ directions. Our analytical results in the limit $k_x \ll 1, k_y \ll 1, \tilde{A}_0\tlda \ll 1, \hbar\Omega \ll t_\sigma$ show the band structure around the $\Gamma$ point are as follows: 
\begin{itemize}
\item{For polarization along the $x$-axis ($\theta=0$), the gap opened at the $\Gamma$ point is $(\tilde{A}_0\tilde{a})^2t_\sigma/16$}.
      Along the $k_x (k_y=0)$ axis, two Dirac points appear at                 
$\pm (\tilde{A}_0\tilde{a}) / \sqrt{6}$, while
      along $k_y (k_x=0)$ a gap is opened.
\item{For polarization along the $y$-axis ($\theta=\pi/2$), the gap opened at the $\Gamma$ point is $7(\tilde{A}_0\tilde{a})^2t_\sigma/48$}. As shown in Fig.\ref{fig:floquet-band-nn-lin-THz}(c-d), no further Dirac points are formed.
\end{itemize}
Figs.\ref{fig:floquet-band-nn-lin}(b-d) show the band structure ($t'/t_\sigma=0.0$) for linearly polarized light 
with fixed amplitude $\tilde{A}_0\tlda =1.0$ and frequency $\hbar\Omega/t_{\sigma}=10$.
We study the dependence of the Floquet-Bloch band structure on the polarization direction of the linear polarized light.
By setting $\theta=\pi/2$, the electric field is applied along the $y$ axis. The band structure is plotted  
along the $x$ direction in Fig.\ref{fig:floquet-band-nn-lin}(b). 
The effect of the $y$-polarized laser is to split the quadratic touching point into two Dirac points on the $k_x$ axis. 
A gap is opened at the Dirac points located at $\bfK (\bfK')$.
% \theta = 0 -- electric field point along x direction 
Next, we set the electric field to be along the $x$-axis ($\theta=0$). The effect on the band structure is shown in 
Figs.\ref{fig:floquet-band-nn-lin}(c-d). 
The Dirac points located at $\bfK (\bfK')$ undergo a tiny shift towards the $\Gamma$ point.
The effect of $x$-polarized laser is to split the quadratic touching point into two Dirac points on the $k_y$ axis. 

Note that the behavior of the quadratic band touching point in linearly polarized light is rather different depending on whether the light is on-resonant or off-resonant.  In particular, Dirac points appear only along the $x$-direction for polarization along the {\it same} direction in the on-resonant case, while in the off-resonant case Dirac points appear in the perpendicular direction for both $x$ and $y$ polarizations.  Thus, the high-frequency regime exhibits more universal behavior than the low-frequency regime. 

We next take the second neighbor hopping into consideration. Because the effect of a linearly polarized laser on the Dirac points is very similar to the previous case ($t'=0$), we focus on the effect of linearly polarized light on the quadratic touching points. 
The Floquet-Bloch band structure with linearly polarized light is shown in Figs.\ref{fig:floquet-band-nnn-lin}(c-f). 
Here we set the polarization of laser to be along the $x$-axis ($\theta=0$) and plot the band structure along the $k_y$-axis.
The amplitude of laser is fixed at $\tilde{A}_0\tlda=\sqrt{2}/10$. In the theoretical infinite frequency limit, the effective Hamiltonian will approximately be the time-averaged time dependent Hamiltonian.\cite{Bukov:ap15,Mikami:prb16} 

Our analytical results in the high frequency limit show the gap opened at the $\Gamma$ point will be $|3(\tilde{A}_0\tlda)^2(t_\sigma-12t')/16|$. Fig.\ref{fig:floquet-band-nnn-lin}(c) shows the band structure in the high frequency limit ($\hbar\Omega/t_\sigma=100$) to mimic the behavior of the theoretical infinite frequency limit. Two new Dirac points are situated at $(0, \pm 0.1)$, which is close to (0, $\pm \tilde{A}_0/\sqrt{2}$). 
Reducing the frequency of laser, the gap at the $\Gamma$ point tends to close and the two Dirac points move toward the $\Gamma$ point, as shown in Fig.\ref{fig:floquet-band-nnn-lin}(d) with $\hbar\Omega/t_\sigma=10$. 
Continuing to decrease the frequency ($\hbar\Omega/t_\sigma=8.33$) will merge the two Dirac 
points into one quadratic touching point at $\Gamma$, as shown in in Fig.\ref{fig:floquet-band-nnn-lin}(e). 
Further decreasing the frequency, the quadratic touching point will open a gap without 
new Dirac points formed.  The gap for $\hbar\Omega/t_\sigma=6.67$ is shown in Fig.\ref{fig:floquet-band-nnn-lin}(f).  Note the inclusion of the second neighbor hopping, $t'$, generally leads to dispersive bands in equilibrium, Fig.\ref{fig:floquet-band-nnn-lin}(a).

\subsection{Circularly Polarized Light}
For circularly polarized light with vector potential $\bfA(t)=A_0[\cos(\Omega t), -\sin(\Omega t)]$, the Floquet Hamiltonian matrix elements are,
\begin{align}
     H^{AB}_{nm}(\bfk) &= [t_x f(k_2,\alpha_2) + t_y f(k_3,\alpha_3) + t_z] \mathcal{J}_{m-n}(\tilde{A}_0\tlda), \nonumber\\
     H^{BA}_{nm}(\bfk) &= [t_x^* f(k_2,\alpha_2) + t_y^* f(k_3,\alpha_3) + t_z^*] \mathcal{J}_{m-n}(-\tilde{A}_0\tlda), \nonumber\\
     H^{AA}_{nm}(\bfk) &= [t_{xy} f(-k_1,\beta_1) + t_{zx} f(-k_2, \beta_2) \nonumber\\ 
                       &\quad\quad\quad\ \ + t_{yz} f(-k_3, \beta_3)] \mathcal{J}_{m-n}(-\sqrt{3}\tilde{A}_0\tlda) \nonumber\\
                       &+ [t_{xy} f(+k_1,\beta_1) + t_{zx} f(+k_2, \beta_2) \nonumber\\ 
                       &\quad\quad\quad\ \ + t_{yz} f(+k_3, \beta_3)] \mathcal{J}_{m-n}(+\sqrt{3}\tilde{A}_0\tlda),
\end{align}
where $f(k_i, x) = e^{i \bfk \cdot a_i} e^{i(m-n)x}$ and $\alpha_2 = -\alpha_3=2\pi/3$, 
$\beta_1=\pi/2, \beta_2 = -\beta_3 = \pi/6$.

Diagonalizing the time-independent Hamiltonian in the truncated Floquet space will give one the Floquet-Bloch band structure.
Fig.\ref{fig:floquet-band-nn-cir-THz} shows the Floquet-Bloch band structure very close to $\Gamma$ point at quarter filling for the case of circular polarization. We plot the band structure as a function of
laser intensity while fixing its frequency at $5$ THz. As the laser intensity is increased, the size of the gap at $\Gamma$ point is increasing monotonically, at the same time, 
the two bands are pushed up correspondingly. This kind of behavior can be understood by deriving the low energy Hamiltonian at $\Gamma$. 
By downfolding the Floquet Hamiltonian in the limit $k_x=k_y=0, \tilde{A}_0\tilde{a} \ll 1$ and keeping terms $\mathcal{O}(\tilde{A}_0^2\tilde{a}^2)$, 
we find that, the bands are pushed up by $3(\tilde{A}_0\tilde{a})^2t_\sigma/16$ and 
the gap opened at $\Gamma$ point at quarter filling is $(\tilde{A}_0\tilde{a})^2\Omega/8$.

Figs.\ref{fig:floquet-band-nn-cir}(a-c) shows the band structure ($t'/t_\sigma=0.0$) for circularly polarized light 
with amplitudes $\tilde{A}_0\tlda=0.5, 1.7, 3.8$ and fixed driving frequency $\hbar\Omega/t_{\sigma}=10$.
The dominant features of the band structure can be understood by taking the effective Hamiltonian in the infinite frequency limit.\cite{Bukov:ap15}
In this limit, the effective Hamiltonian will be 
\begin{equation}
H_{\mathrm{eff}}(\bfk) = \frac{1}{T} \int_{0}^{T} H(\bfk, t) dt = \mathcal{J}_0(\tilde{A}_0\tlda) H_0(\bfk, t'=0),\nonumber
\end{equation}
which means the original bands in equilibrium are renormalized by a scale factor of the zero-$th$ order Bessel function. 
The photon-dressed band structure in infinite frequency limit are shown as dashed lines in Fig.\ref{fig:floquet-band-nn-cir}.
As a result, in this limit, the laser will rescale the bands by $\mathcal{J}_0(\tilde{A}_0\tlda)$, which could even be zero or negative. The 
For example, $\mathcal{J}_0(0.5)=0.938$ in Fig.\ref{fig:floquet-band-nn-cir}(a). 
Continuing to increase the amplitude of the laser will tend to renormalize bands towards zero bandwidth, as shown in Fig.\ref{fig:floquet-band-nn-cir}(b). 
Finally, when the Bessel function changes sign, the bands will be scaled by $|\mathcal{J}_0(\tilde{A}_0\tlda)|$ and be inverted ($\mathcal{J}_0(3.8)=-0.403$) 
in Fig.\ref{fig:floquet-band-nn-cir}(c). By comparing the numerical exact Floquet-Bloch band structure with the photon-dressed band structure in infinite frequency limit, we verified that
the zeroth order approximation in Floquet-Magus expansion plays a dominant role 
throughout the parameter range used, while high-order correction plays an important role in determining the detailed band structure, for example, bands around quadratic touching points shown in inset of Fig.\ref{fig:floquet-band-nn-cir}(a). 

Adding the next-nearest neighbor terms ($t'/t_\sigma=0.1$) will make the results considerably more interesting.
Figs.\ref{fig:floquet-band-nnn-cir}(a-f) shows the band structure ($t'/t_\sigma=0.1$) for circularly polarized light 
with amplitudes $\tilde{A}_0\tlda=0.50, 1.39, 1.80, 2.40, 2.83, 3.80$ and fixed driving frequency $\hbar\Omega/t_{\sigma}=10$.
To understand the Floquet-Bloch band structure shown in Figs.\ref{fig:floquet-band-nnn-cir}, 
we take the theoretical infinite frequency limit as before, the effective Hamiltonian will be 
\begin{align}
    &H_{\mathrm{eff}}(\bfk) = \frac{1}{T} \int_{0}^{T} H(\bfk, t) dt \nonumber\\
    &= \mathcal{J}_0(\tilde{A}_0\tlda) H_0(\bfk, t_\sigma, t'=0) + \mathcal{J}_0(\sqrt{3} \tilde{A}_0\tlda) H_0(\bfk, t_\sigma=0, t') \nonumber\\
    &= H_0(\bfk, t'=0, t^\sigma_{\mathrm{eff}}) +  H_0(\bfk, t^\sigma=0, t_{\mathrm{eff}}'), \nonumber
\end{align}
where we defined $t^\sigma_{\mathrm{eff}}=\mathcal{J}_0(\tilde{A}_0\tlda)t_\sigma$ and $t_{\mathrm{eff}}'=\mathcal{J}_0(\sqrt{3} \tilde{A}_0\tlda)t'$.
This effective Hamiltonian tells us the original nearest and next-nearest neighbor hopping terms in equilibrium are renormalized 
by a scale factor of the zero-$th$ order Bessel function $\mathcal{J}_{0}(\tilde{A}_0\tlda)$ and $\mathcal{J}_0(\sqrt{3}\tilde{A}_0\tlda)$, respectively.
We already know that, in the absence of the laser, the effect of the next nearest neighbor hopping terms $t'$ is to break the particle-hole symmetry. 

As one switches on the laser with small amplitude $\tilde{A}_0\tlda=0.5$ (for $0<\tilde{A}_0\tlda<1.39$, we have $\tpeff < 0.1\tseff$), 
an effect of the circularly polarized laser is to increasingly recover the particle-hole symmetry, as shown in Fig.\ref{fig:floquet-band-nnn-cir}(a). 
At $\tilde{A}_0\tlda=1.39$ ($\tpeff=0$), the particle-hole symmetry 
is fully recovered, shown in Fig.\ref{fig:floquet-band-nnn-cir}(b).
For $1.39<\tilde{A}_0\tlda<2.404$, the effective hopping parameters are $\tseff > 0$ and $\tpeff < 0$, which push the bands upward, shown in Fig.\ref{fig:floquet-band-nnn-cir}(c).
Around $\tilde{A}_0\tlda=2.404$, $\tseff \approx 0, \tpeff \approx -0.3812$, the next nearest neighbor term is dominant in the Floquet-Bloch band structure, as shown in Fig.\ref{fig:floquet-band-nnn-cir}(d).
At $\tilde{A}_0\tlda=2.83$, $\tpeff =0.1\tseff <0$, the band structure is exactly a band inversion, as shown in Fig.\ref{fig:floquet-band-nnn-cir}(e).
Around $\tilde{A}_0\tlda=3.800$, we have $\tseff<0$ and $\tpeff >0$, and the bands are as shown in Fig.\ref{fig:floquet-band-nnn-cir}(f).

\subsection{Comparison with results for the nickel-oxygen tight-binding model}
The nickel-oxygen model\cite{Ruegg:prb12} is written as
\begin{align}
     H_{\mathrm{Ni-O}} = 
     \sum_{i\sigma} \epsilon_{p} p_{i\sigma}^\dagger p_{i\sigma} +
     \sum_{i\alpha\sigma} \epsilon_{d} d_{i\alpha\sigma}^\dagger d_{i\alpha\sigma} +
     H_{\mathrm{hyb}} + H_{p-p},
     \label{eq:ni-ox-tb}
\end{align}
where $p_{i\sigma}^\dagger (p_{i\sigma})$ creates (annihilates) an electron in an oxygen $p$-orbital with spin $\sigma$, $d_{i\alpha\sigma}^\dagger (d_{i\alpha\sigma})$ creates (annihilates) an electron in the nickel $e_g$ orbital with band $\alpha$ and spin $\sigma$. $H_{\mathrm{hyb}}$ describes the hybridization between oxygen $p$-orbital electrons and Ni $e_g$-orbital electrons.  The hybridization is parametrized by the Slater-Koster parameter $V_{pd\sigma}=1.8$eV. $H_p$ describe the hopping between oxygen $p$-orbitals parameterized by Slater-Koster parameter $V_{pp\sigma}=1.4$eV. Here we follow the Ref.[\onlinecite{Ruegg:prb12}], and choose the parameters as $\epsilon_p^{in} = -4.74 $eV, $\epsilon_p^{out} = -5.47 $eV, $\epsilon_d = -1.47 $eV. Here we set the in-layer (sandwiched by Ni-Ni) and out-layer (sandwiched by Ni-Al) oxygen ions to have different on-site potential energy.

\begin{figure}[th]
\begin{center}
\includegraphics[width=4cm]{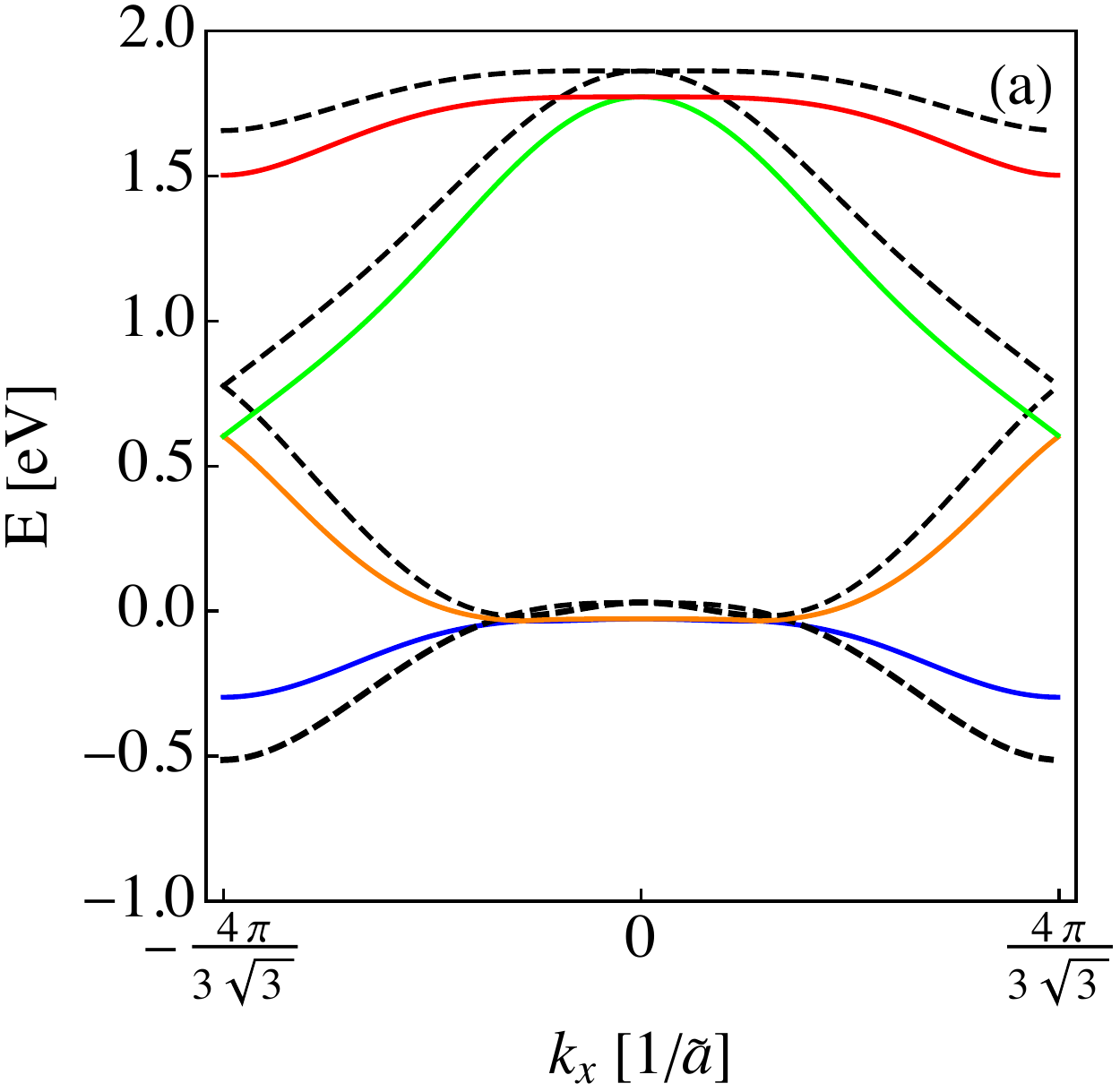}  %
\includegraphics[width=4cm]{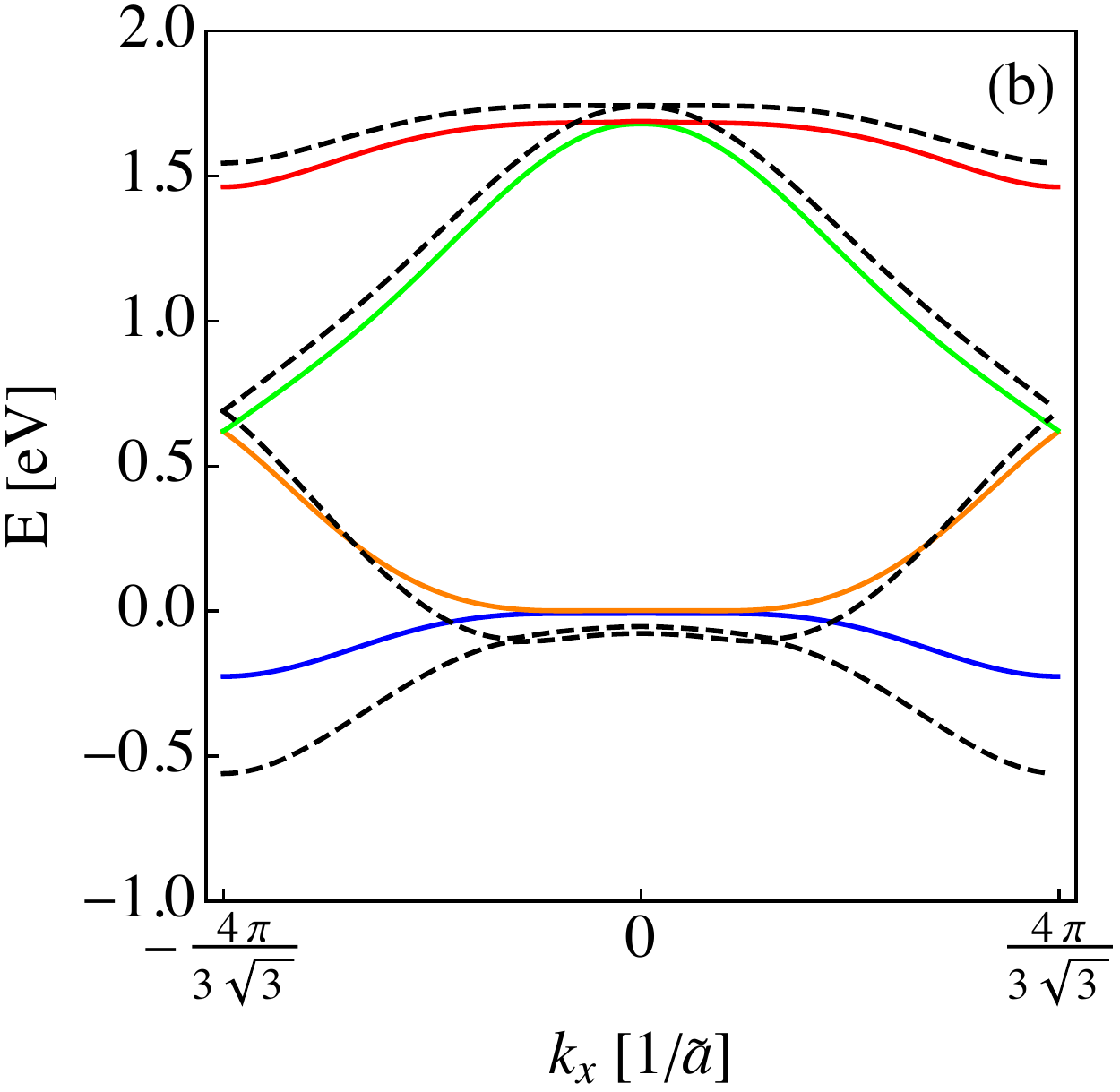}  %
\includegraphics[width=4cm]{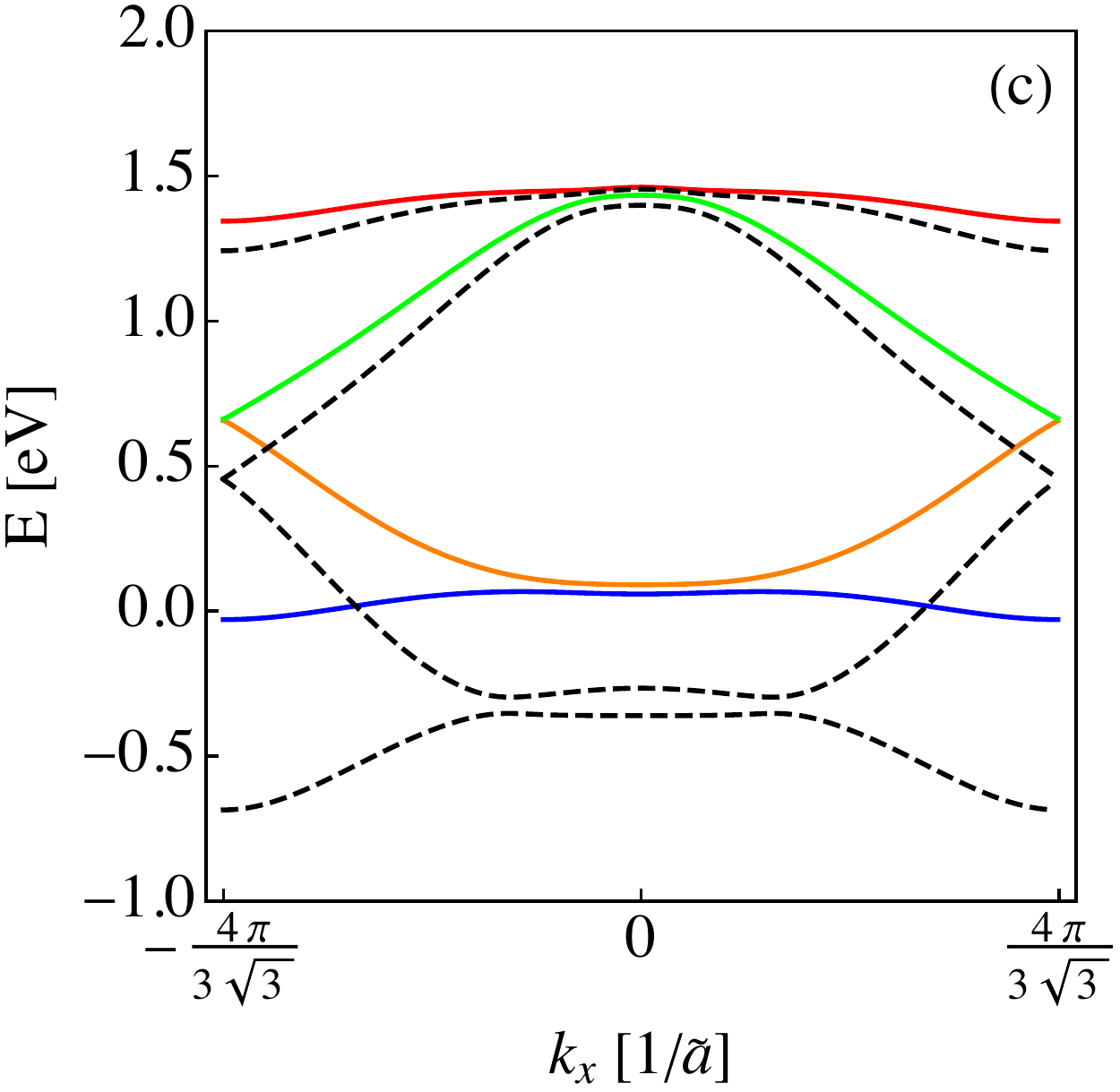}  %
\includegraphics[width=4cm]{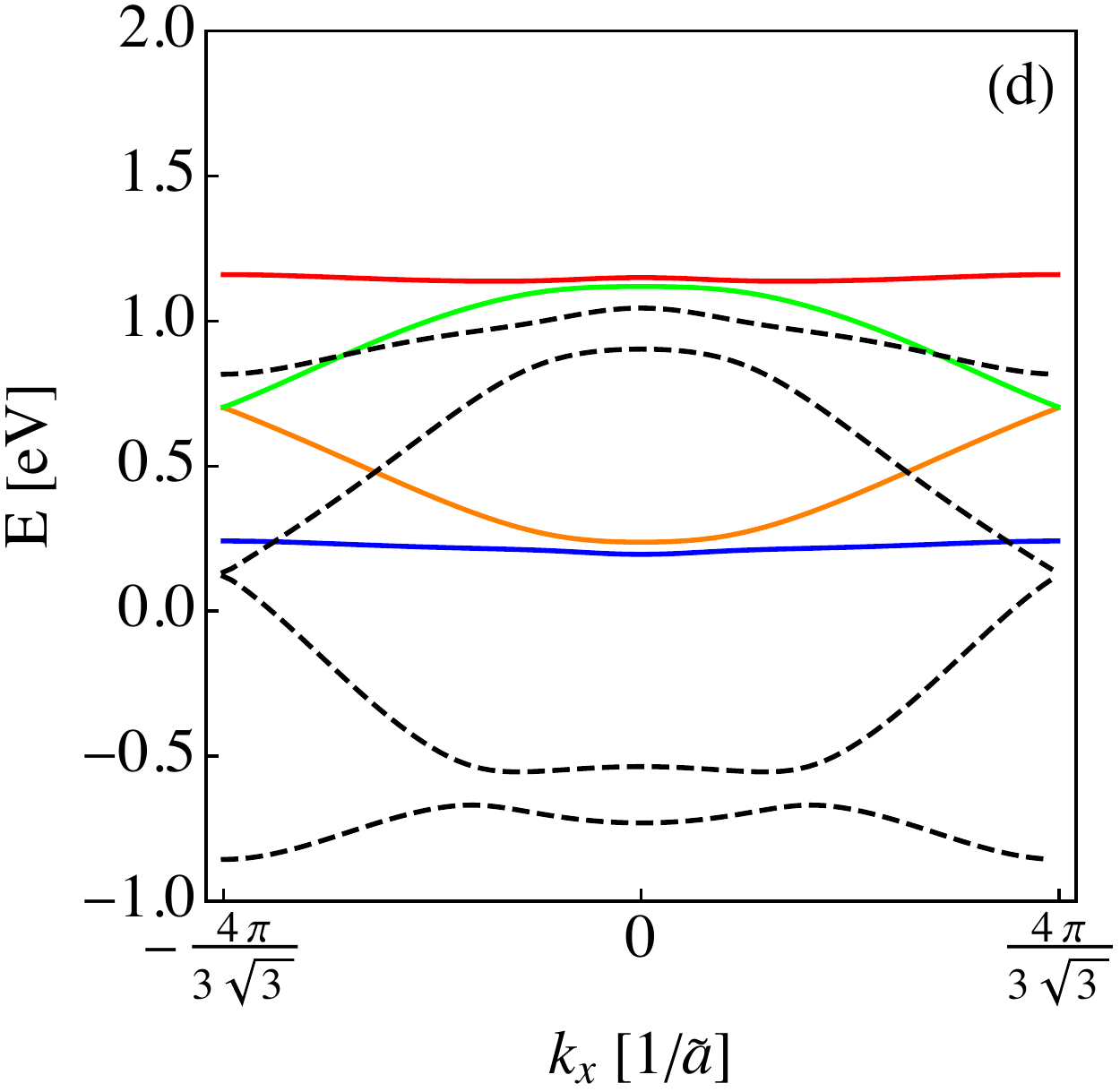}  %
\includegraphics[width=4cm]{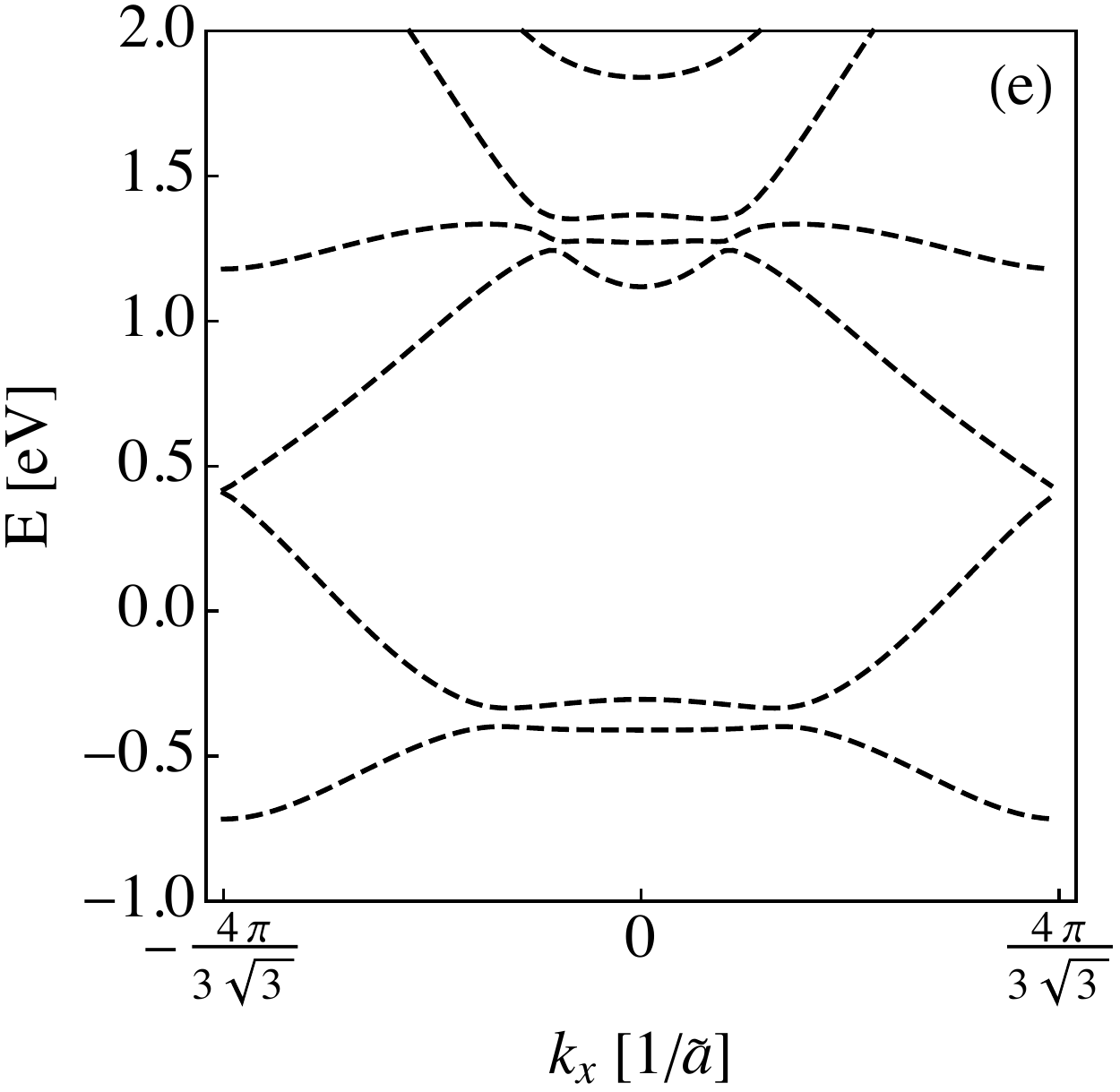}  %
\includegraphics[width=4cm]{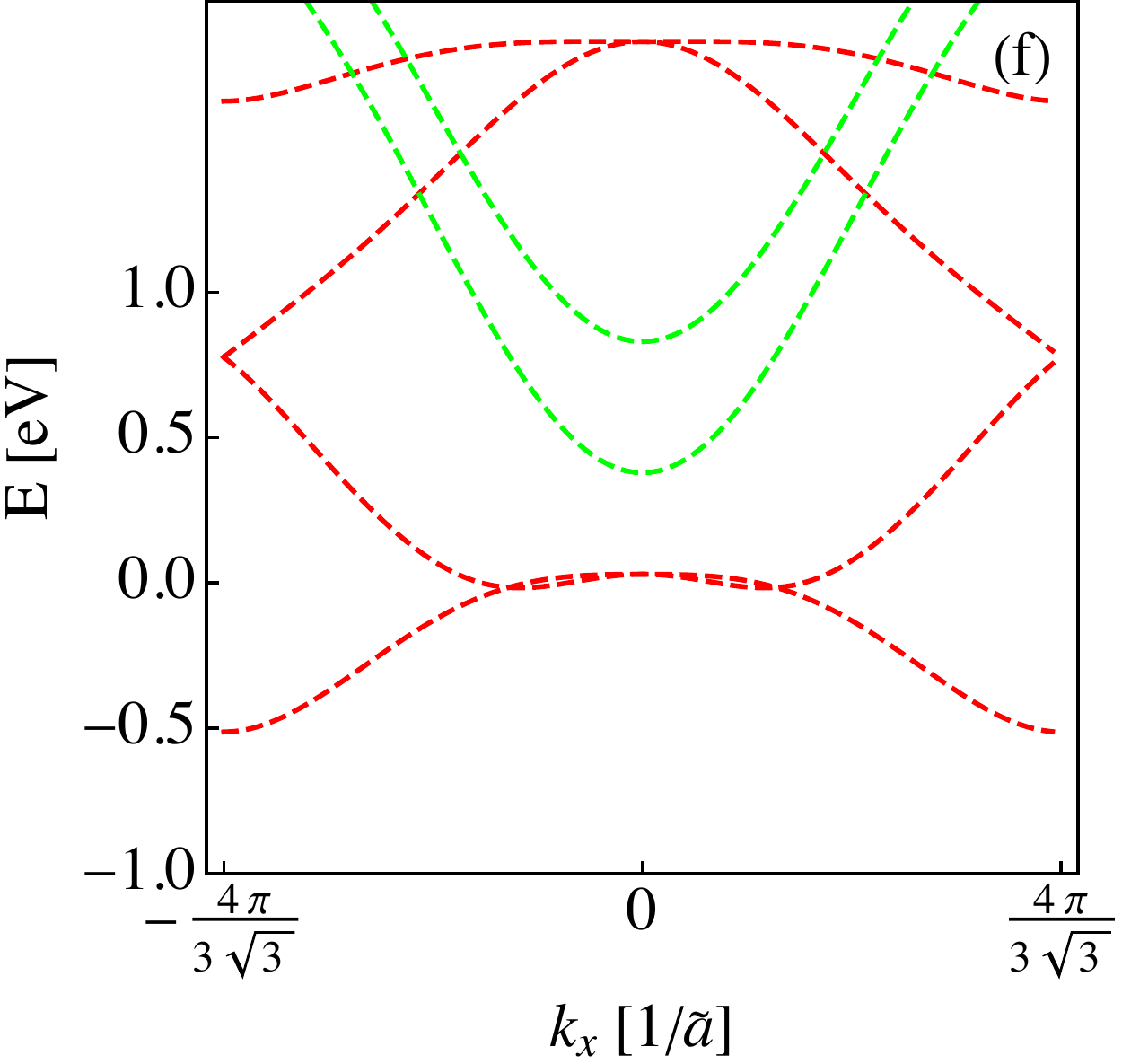}  %
\end{center}
\caption{(Color online) The Floquet-Bloch band structure of the (111) bilayer LaNiO$_3$ embedded in a normally incident circularly polarized light
$\bfA(t) = A_0(\cos\Omega t, -\sin\Omega t)$ along the $k_x (k_y=0)$ direction.
Both nearest neighbor and next nearest neighbor hopping terms are kept in the tight-binding Hamiltonian Eq.\eqref{eq:htbk-t} 
with $t_\sigma=0.6$eV, $t'=0.1t_\sigma$.
The band structure are shown with solid lines for Nickel-Nickel tight-binding model in Eq.\eqref{eq:ni-ni-tb}, dashed lines for Nickel-Oxygen tight-binding model in Eq.\eqref{eq:ni-ox-tb}.
(a) Equilibrium (absence of laser) band structure;
(b) $\tilde{A}_0 \tlda=0.50$, $\hbar\Omega=12$ eV;
(c) $\tilde{A}_0 \tlda=1.00$, $\hbar\Omega=12$ eV;
(d) $\tilde{A}_0 \tlda=1.50$, $\hbar\Omega=12$ eV;
(e) $\tilde{A}_0 \tlda=1.00$, $\hbar\Omega=10.5$ eV;
(f) $\tilde{A}_0 \tlda=0.00$ and $\hbar\Omega=10.5$ eV. (f) and (e) only differ in that there is no coupling between Floquet states in (e).
}
\label{fig:floquet-band-nio}
\end{figure}

In Fig.\ref{fig:floquet-band-nio}, we plot the Floquet-Bloch band structure for the Ni-Ni tight-binding model in Eq.\eqref{eq:ni-ni-tb} and Ni-O tight-binding model in Eq.\eqref{eq:ni-ox-tb} with solid and dashed lines, respectively.
Fig.\ref{fig:floquet-band-nio}(a) shows the band structure in equilibrium (absence of laser). The total bandwidth of the Ni-O model is around 12 eV. For a comparison of the Floquet-Bloch band structure for the two tight-binding models, we plot the band structure in Fig.\ref{fig:floquet-band-nio}(b-d) with the amplitude $\tilde{A}_0 \tlda=0.5,1.0,1.5$ and frequency $\hbar\Omega=12$ eV.
Note that the band structure is scaled differently with frequency in the Ni-Ni model and the Ni-O model due to greater proximity to resonant transitions in the Ni-O model with a wider overall bandwidth. This highlights the importance of higher energy bands in general Floquet situations and is relevant to experimental efforts to realize Floquet topological insulators, and other desired Floquet band structures.

Finally, we consider the resonant process between O $p$-bands and Ni $e_g$-bands by setting the laser frequency to be $\hbar\Omega=10.5$ eV, which is less than the bandwidth of the Ni-O model. The equilibrium bands (red) and Floquet band copies of oxygen $p$-bands (green) are plotted in Fig.\ref{fig:floquet-band-nio}(f). The Floquet-Bloch bands are shown in Fig.\ref{fig:floquet-band-nio}(e). Due to the overlap between the oxygen $p$ and nickel $e_g$ bands, band inversion occurs and a gap opens at the crossing points.  Hence, in a multi-band model is the total bandwidth that is very important for the band Floquet band structure for resonant conditions. 
% figures with both nn and nnn hopping terms

\section{Low energy effective Hamiltonian for quadratic band touchings and Dirac points}
\label{sec:lowenergy}
In the previous section, we saw that the dominant band structure features of the Floquet-Bloch bands for large driving frequencies can be understood by the time averaged Hamiltonian.  However, to see how the quadratic band touching points and Dirac band touching points respond to the circularly and linearly polarized light, it is helpful to derive an effective low-energy theory to better understand the universal features.  Because the Dirac band touching point has been much discussed in the literature,\cite{Inoue:prl10,Oka:prb09,Kitagawa:prb11,Iadecola:prl13,Ezawa:prl13,Gu:prl11,Hossein:prb14,Aditiprb91-2015,Aditiprb92-2015,Hossein:prb16} here we focus on the quadratic band touching point at high frequency, which to the best of our knowledge has received very little attention.\cite{Du:prb2017}

The effective Hamiltonian describing the quadratic band touching point at $1/4$ filling (in the absence of the drive, and in the presence of time-reversal symmetry) can be written in a general quadratic band touching form\cite{Sun:prl09} (in the absence of time-reversal symmetry breaking\cite{Balazs:prb14}),
\begin{equation}\label{eq:qbc_gen}
    H_{\qbc}(k) = \eta_I (k_x^2 + k_y^2) \mathbb{I} + 
                2 \eta_x k_x k_y \sigma_x + \eta_z (k_x^2 - k_y^2) \sigma_z,
\end{equation}
where $\mathbb{I}$ is the identity matrix, and $\sigma_x$ and $\sigma_z$ are two real Pauli matrices along $x$ and $z$, respectively.
The three coefficients above are determined as, 
\begin{eqnarray}
\eta_I&=&3(t_\sigma-12t'+t_\delta)/16 + 3t_\sigma t_\delta/(4t_\sigma+4t_\delta),\nonumber \\
\eta_z&=&3(t_\sigma-12t'-t_\delta)/16,\nonumber \\
\eta_x&=&-\eta_z.
\end{eqnarray}
By setting $t_\delta=0$, the effective Hamiltonian simplifies to
\begin{equation}\label{eq:qbc}
    H_{\qbc} = \frac{3}{8}(t_\sigma-12t')
    \begin{pmatrix}
        k_x^2  & - k_x k_y \\
       -k_x k_y   & k_y^2
    \end{pmatrix}.
\end{equation}
Technical details related to the derivation of the effective Hamiltonian can be found in Ref.[\onlinecite{Ruegg_Top:prb13}] 
using second order perturbation theory or Ref.[\onlinecite{Du:prb2017}] using downfolding. In the remaining part of this 
paper, all the effective two-band Hamiltonians are derived using downfolding. 

In the following, we derive the low energy effective two band Hamiltonian up to second order in $k_x\tlda \ll 1, k_y\tlda \ll 1$, and $\tilde{A}_0\tlda \ll 1$.
For this specific model Hamiltonian, 
the effective Hamiltonian derived from the Floquet Hamiltonian is {\em no longer} the one obtained by 
setting $\bfk \rightarrow \bfk + e\bfA(t)/\hbar$ in Eq.\eqref{eq:qbc}. 
Instead, one needs to follow the two-step precedure detailed in Ref.[\onlinecite{Du:prb2017}] to derive a $2 \times 2$ effective Hamiltonian.  The results of the low-energy theories are summarized in Table~\ref{table:low_energy_gap}.

\begin{table}[h]
\centering
\begin{tabular}{|c|c|c|}
\hline 
Light Polarization & Gap on-resonance & Gap off-resonance \\
\hline 
circular & $(\tilde A_0 \tilde a)^2 (\hbar\Omega)/8$ & 9 $(\tilde A_0 \tilde a)^2t_\sigma/(8\hbar\Omega)$\\
\hline
linear, $\theta=0$ & $(\tilde A_0 \tilde a)^2 t_\sigma/16$ & $3 (\tilde A_0 \tilde a)^2  t_\sigma/16$ \\
\hline
linear, $\theta=\pi/2$ & $7(\tilde A_0 \tilde a)^2 t_\sigma/48$ & $3 (\tilde A_0 \tilde a)^2  t_\sigma/16$\\
\hline
\end{tabular}
\caption{\label{table:low_energy_gap} Gap dependence at the $\Gamma$ point for 1/4 band filling as a function of light polarization and energy of drive.  On-resonant drives have small photon energy compared to the bandwidth and off-resonant drives have large photon energies compared to bandwidth.}
\end{table}

\subsection{Linearly Polarized Light}
\label{low-lin}
We have seen that linearly polarized light along the $x$-axis in the plane can be expressed as $\bfA(t) = A_0 \cos(\Omega t) (1, 0)$. 
The effective Hamiltonian around the quadratic touching point at $1/4$ filling is given by,
\begin{align}\label{eq:qbc-lin}
    &H_{\qbc} = \frac{3}{8}(t_\sigma-12t')
    \begin{pmatrix}
        k_x^2  & - k_x k_y \\
       -k_x k_y   & k_y^2
    \end{pmatrix} \nonumber\\
    + &\frac{3(\tilde{A}_0\tlda)^2}{32}
    \begin{pmatrix}
        3 (t_\sigma-8t') + \frac{27t_\sigma^3}{(\hbar\Omega)^2}  & 0 \\
        0  & t_\sigma + \frac{3t_\sigma^3}{(\hbar\Omega)^2}
    \end{pmatrix}.
\end{align}
The magnitude of the gap at the ${\bf \Gamma}$ point can be obtained from the low-energy form of the Hamiltonian, Eq.\eqref{eq:qbc-lin}, and is $|3(\tilde{A}_0\tlda)^2(t^\sigma-12t')/16+9(\tilde{A}_0\tlda)^2t_\sigma^3/4(\hbar\Omega)^2|$.
The gap magnitude will be simplified to $|3(\tilde{A}_0\tlda)^2(t_\sigma-12t')/16|$ in the high frequency limit. Decreasing the laser frequency will result in a reduced gap size (since $t^\sigma<12t'$).

Further, by diagonalizing the low energy Hamiltonian in Eq.\eqref{eq:qbc-lin}, we realize that two conditions need to be simultaneously 
fulfilled to open a gap and prevent new Dirac points from being formed: (1) $t_\sigma-12t' < 0$, (2) $(\hbar\Omega)^2 > 12t_\sigma^3/(12t'-t_\sigma)$. 
Otherwise, the quadratic touching point will split into two Dirac points with the position of two new Dirac points being $\pm (0, \sqrt{1-12t_\sigma^3/[(12t'-t_\sigma)(\hbar\Omega)^2]})\tilde{A}_0/\sqrt{2}$.

In the high frequency limit, the position of the two Dirac points are $\pm(0,\tilde{A}_0/\sqrt{2})$. Decreasing the frequency will pull the two Dirac points toward the $\Gamma$  point. There exist a critical frequency $\hbar\Omega_c=\sqrt{12 t_\sigma^3/(12t'-t_\sigma)}$ where the two Dirac points will merge into one quadratic touching point at the $\Gamma$ point. Further decreasing the laser frequency, $\Omega < \Omega_c$, the quadratic touching points will open a gap and no new Dirac points are formed.

We also study the effect of a linearly polarized laser on the quadratic touching point at $3/4$ filling. 
It will split into two Dirac points situated at $\sqrt{1+12 t_\sigma^3/[(t+12t')(\hbar\Omega)^2]}\tilde{A}_0/\sqrt{2}$ which 
shift the position for infinity frequency, $\tilde{A}_0/\sqrt{2}$, away from $\Gamma$ point.

One can contrast the behavior of the quadratic band touching point with what happens at the Dirac points.  
By keeping terms up to quadratic order in momentum (leading corrections to the pure 
Dirac dispersion), one can derive the low energy effective Hamiltonian around the 
Dirac points at $\mathbf{K}$ ($\mathbf{K'}$). We find that the 
response of the Dirac point depends on the polarization direction of the pump light: For $\theta=\pi/2$ 
(and symmetry related directions), the middle two bands will open a gap. 
Away from $\theta = \pi/2$ (and symmetry related directions), the Dirac points will undergo a small shift and remain gapless.
These results apply to Dirac points at both $\mathbf{K}$ and $\mathbf{K}'$ points.

\subsection{Circularly Polarized Light}
\label{low-cir}
The effect of circularly polarized light on the quadratic band touching point is rather different from the case of linearly polarized light. 
The effective Hamiltonian around quadratic touching point at $1/4$ filling is given by,
\begin{align}\label{eq:qbc-cir}
    H_{\qbc} = &\frac{3}{8}(t_\sigma-12t')
    \begin{pmatrix}
        k_x^2  & - k_x k_y \\
       -k_x k_y   & k_y^2
    \end{pmatrix} \nonumber\\
    + &\frac{3(\tilde{A}_0\tlda)^2}{16}
    \begin{pmatrix}
        2 (t_\sigma-6t')  & -3it_\sigma^2/\hbar\Omega \\
         3it_\sigma^2/\hbar\Omega  &  2(t_\sigma-6t')
    \end{pmatrix}.
\end{align}
%Taking the high frequency limit, the effective 
%Hamiltonian shows that the band will shift by $2A_0^2(t_\sigma-6t')$ and no gap at $\Gamma$ is opened.
The magnitude of the gap at the ${\bf \Gamma}$ point can be obtained from the low-energy form of the Hamiltonian, Eq.\eqref{eq:qbc-cir}, and is $9(\tilde{A}_0\tlda)^2 t_\sigma^2/(8\Omega)$. The gap will be close in the high frequency limit.
By contrast, the gap at the Dirac points $\mathbf{K}$ and $\mathbf{K'}$ are proportional to 
$9\tilde{A}_0^2t_\sigma t_\delta/(2\hbar\Omega) + 81 \tilde{A}_0^2 t'^2/(2\hbar\Omega) + 81\tilde{A}_0^2(t_\sigma^2+6t'^2)^2/[32(\hbar\Omega)^3]$, which also closes in the high-frequency limit but as a different power of $\hbar\Omega$ if $t_\delta=t'=0$.

\section{OPTICAL HALL CONDUCTIVITY}
\label{sec:optical}
One way the Floquet-Bloch band features discussed above can be probed in experiments is via the optical Hall conductivity. The finite-frequency transverse optical conductivity--related to the Faraday rotation\cite{Aditiprb92-2015,Morimoto:prl09,O'Connell:prb82,Ikebe:prl10,Crassee:np11,Shimano:nc13}--can reveal some of the features in the band structure, particularly at energies where transitions occur between ``flat" regions (in momentum space) of the bands.  Typically,  transport measurements on Floquet systems are problematic\cite{Farrell:prb16,Torres:prl14,Rudner:prx13,Kitagawa:prb11}  and angle resolved photoemission spectroscopy (ARPES),\cite{Wang:sci13,Mahmood:np16} only provides access to the occupied states.  

If one defines a Fourier transformed Berry ``vector potential" as,\cite{Aditiprb91-2015}
\begin{equation}
     A_{\beta i \alpha}^m = \frac{1}{T} \int_0^T dt e^{-i m\Omega t} \langle\phi_{k\beta}(t)|\frac{\partial}{\partial k_i}\phi_{k\alpha}(t)\rangle,
\end{equation}
then one can express a quantity $F_k^m$ in terms of them as
\begin{equation}
\label{eq:Fkm}
     F_k^{m,\alpha\beta} = i\left[A_{\beta x \alpha}^{-m} A_{\alpha y \beta}^m - A_{\beta y \alpha}^{-m} A_{\alpha x \beta}^m \right],
\end{equation}
which leads to a linear response formula\cite{Aditiprb91-2015} for the optical conductivity in terms of the time-averaged Berry curvature of a period of the drive, 
\begin{equation}
     \bar{F}_{k \alpha} = \frac{1}{T} \int_0^T 2 \Im [\langle\partial_y \phi_{k\alpha}(t)|\partial_x \phi_{k\alpha}(t)\rangle] dt,
\end{equation}
where $\alpha$ is the band index, $ \Im $ denotes the imaginary part, and  $\bar{F}_{k\alpha} = \sum_{\beta, m} F_{k}^{m,\alpha\beta}$.
Technical details of the numerical calculation of $F_k^m$ can be found in the appendix of Ref. [\onlinecite{Du:prb2017}].

For a four-band system, the optical Hall conductance\cite{Aditiprb91-2015,Aditiprb92-2015} can be written as a sum,
\begin{align}
    \sigma_{xy}(\omega) = \sum_{\alpha<\beta}\sigma_{xy}^{\alpha\beta}(\omega)
                        = \sum_{m=\mathrm{int}}\sum_{\alpha<\beta}\sigma_{xy}^{\alpha\beta}(\omega),
\end{align}
with
\begin{align}
\label{eq:hall-m}
     \sigma_{xy}^{m,\alpha\beta}(\omega) &= -\frac{e^2}{2\pi h} \int d^2 k E_{m,\alpha\beta}^2 F_k^{m,\alpha\beta} \nonumber\\
                 &\times \frac{\omega^2 - E_{m, \alpha\beta}^2 - 2i\omega\delta}
                 {[\omega^2 - E_{m,\alpha\beta}^2]^2 + 4\omega^2\delta^2}[\rho_{k\alpha} - \rho_{k\beta}],
\end{align}
where
\begin{equation}
     E_{m,\alpha\beta} = E_{k\beta} - E_{k\alpha} - m\Omega.
\end{equation}
The terms $\rho_{k\alpha} - \rho_{k\beta}$ in Eq.\eqref{eq:hall-m} provide information about the relative occupations at wave vector $k$ of bands $\alpha$ and $\beta$.    We assume that the system is initially in the ground state of the (LaNiO$_3$)$_2$/(LaAlO$_3$)$_N$ at $1/4$ filling. 
We time evolve the system using $H(t)$, Eq.\eqref{eq:htbk} modified by the vector potential $\bfA(t)$ 
[see Eq.\eqref{eq:htbk-t}] that drives the system into a Floquet-Bloch state.  We consider only the case of circularly polarized light as transverse optical conductivity vanishes in the case of linearly polarized light.\cite{Du:prb2017}

\begin{figure}[t]
\includegraphics[width=1.0\linewidth]{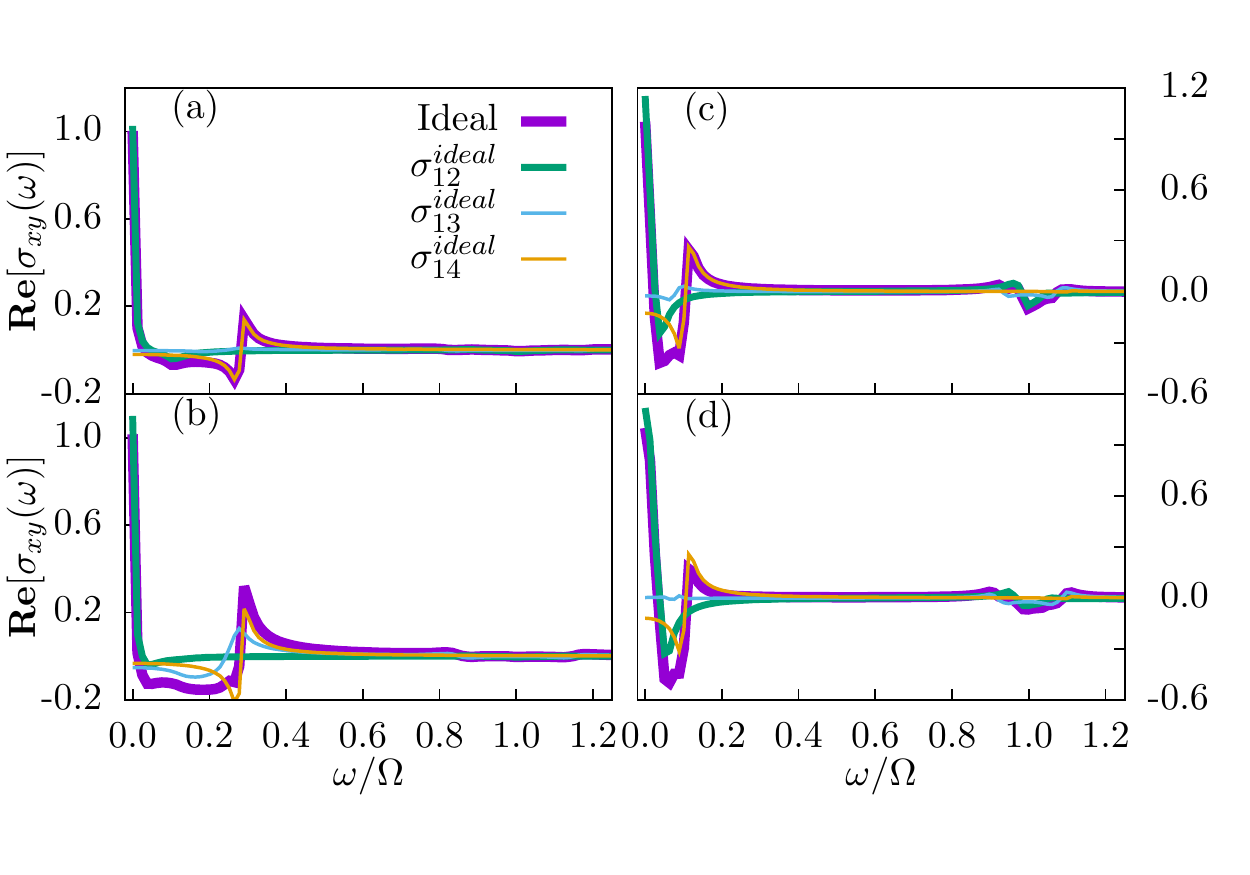}  %
\caption{(Color online) 
Optical Hall conductivity as a function the probe laser frequency. The frequency of the pump laser is fixed to be $\hbar\Omega/t_{\sigma} = 10$.
(a) Only the nearest neighbor hopping is taken into account, $t_{\sigma}=0.6$eV, $t'=0$.
The amplitude of pump laser is set to be $\tilde{A}_0\tlda=0.5$.
(b) Only the nearest neighbor hopping is taken into account, $t_{\sigma}=0.6$eV, $t'=0$.
The amplitude of pump laser is set to be $\tilde{A}_0\tlda=1.8$.
(c) Both the nearest and next-nearest neighbor hopping are taken into account, $t_{\sigma}=0.6$eV, $t'=0.1t_\sigma$.
The amplitude of pump laser is set to be $\tilde{A}_0\tlda=0.5$.
(d) Both the nearest and next-nearest neighbor hopping are taken into account, $t_{\sigma}=0.6$eV, $t'=0.1t_\sigma$.
The amplitude of pump laser is set to be $\tilde{A}_0\tlda=1.8$.
}
\label{fig:optical_hall}
\end{figure}

We refer to the ``ideal" case as when the lowest Floquet-Bloch band of the system is fully occupied while the other two bands are empty (for all time).\cite{Aditiprb91-2015} 
%For the quenched case, the occupation is defined as the overlap between the initial state ($t<0$) and the Floquet quasimode,
%\begin{equation}
 %    \rho_{k\alpha}^{\text{quench}} = |\langle\phi_{k\alpha}(0)|\Psi_{in,k}\rangle|^2,
%\end{equation}
%where $|\Psi_{in,k}\rangle$ is the ground state wave function of the kagome lattice at 1/3 filling  before the laser is switched on at time zero.
Fig.\ref{fig:optical_hall} shows the optical Hall conductivity as a function of the frequency of {\em probe} light.  In the low frequency limit,
the dc Hall conductivity is proportional to the Chern number, $C$: $\sigma_{xy}^{ideal}(\omega=0) = C e^2 / h$ [see Eq.\eqref{eq:hall-m}], and is determined by the $\sum_{m}F_{k}^m$ in Eq.\eqref{eq:Fkm}. Here we used 7 copies $m=-3,\cdots,3$ in the Floquet Hamiltonian for calculations of optical Hall conductivity.  

Clearly,  $\sigma_{xy}^{ideal}(\omega \to 0) =  e^2 / h$ in all cases, so that the circularly polarized light generates topological bands.  Away from the low frequency limit, the individual $F_k^m$ control the behavior.  One can see from Eq.\eqref{eq:hall-m} that the peaks in the integrand are around $\omega \approx |\epsilon_{k\beta} - \epsilon_{k\alpha} - m\Omega|$, while the dominant $k$ are determined by the peaks in $F_k^m$.  The most striking feature that occurs around $\omega/\Omega \approx 0.3$ in Figs.\ref{fig:optical_hall}(a,b) and around $\omega/\Omega \approx 0.18$  Figs.\ref{fig:optical_hall}(c,d) is related to the transitions between the ``lower" and ``upper" most Floquet bands that are relatively weakly dispersing relative to the ``middle energy" bands.  The feature seen around $\omega/\Omega \approx 1.0$  Figs.\ref{fig:optical_hall}(c,d) is related to transitions between ``copies" of the Floquet bands differing by one photon.  The high frequency behavior is $\sigma_{xy}(\omega) \propto 1/\omega^2$.  Thus, the smaller probe frequencies $\omega$ provide the most useful information in experiment.  In addition, it is clear that a relatively small number of transitions dominate $\sigma_{xy}^{ideal}(\omega)$, which can potentially be exploited in experiments or applications that are seeking to be sensitive to a particular frequency of light and relatively insensitive to others.  This insight could be useful in searching for materials that might be good candidates for observing Floquet-Bloch states in the optical conductivity.

\section{DISCUSSION AND CONCLUSIONS}
\label{sec:discussion}
In this work, we theoretically studied the Floquet-Bloch band structure of the bilayer (LaNiO$_3$)$_2$/(LaAlO$_3$)$_N$ heterostructure grown along the (111) direction. We studied the effect of a circularly and linearly polarized laser on the electronic structure using Floquet theory.
In the absence of a laser, first principles calculations suggest that there exists two quadratic touching points 
and two inequivalent Dirac points located at the center ($\Gamma$) and the corners ($K$ and $K'$) of the hexagonal first Brillouin zone,
and two nearly flat bands located at the top and bottom of the electric band structure derived from the $d$-band $e_g$ orbitals closest to the Fermi energy.  A tight-binding fit of the LDA band structure shows that nearest-neighbor hopping plays a dominant role in the band structure, while the next-nearest neighbor hopping breaks the particle-hole symmetry of the system and determines the finer features of the band structure.  An important result that we highlight in this work is hopping terms of different range are renormalized differently under an intensity variation of the light when the driving frequency is held fixed.\cite{Mohan:prb16}  This feature can be used to gain considerable control over the Floquet-Bloch band structure of {\em any} system with hopping parameters of varying distance.  The greater the number of important hopping parameters, the greater the potential control over the Floquet-Bloch band structure.

For the system we studied here, only the first and second neighbor hopping terms play a critical role in the band structure. We found that circularly polarized light can reduce the effect of next-nearest neighbor hopping to be zero while  re-normalizing the effect of nearest neighbor hopping to be finite, resulting in the recovery of particle-hole symmetry, for example. Or, in contrast, the effective nearest neighbor hopping can be renormalized to be zero, rendering the second nearest neighbor hopping terms the dominant ones for the electronic band structure.  

We also introduced a model that explicitly included the oxygen $p$-orbitals and studied the Floquet states of this system. The oxygen $p$-orbitals significantly increased the overall bandwidth of the system and increased the minimum energy of photons for one to be in the off-resonant regime.   The model also highlights the property that the nickel only model and the nickel-oxygen model lead to different results even in the infinite frequency limit, unless the intensity of the light is small, in which case they agree.  Therefore, universal results are only obtained in the large frequency, low-intensity limit.  When trying to interpret experimental systems, it should be born in mind that increasing the intensity of the laser increases the importance of higher energy bands in the system.

With respect to band touching points, we found that there is a gap opened with magnitude $(\tilde{A}_0\tlda)^2/(\hbar\Omega)$ originating in 
1-photon absorption and emission processes at the quadratic touching point (1/4 filling), which is different from a previous study on the kagome lattice (which found a 2-photon process with magnitude $(\tilde{A}_0\tlda)^4/(\hbar\Omega)$). Therefore, the higher energy bands can influence the physics of the gap opening at band touching points. By deriving an effective 2-band Hamiltonian up to second order in momentum using a down-folding scheme, we understand this happens due to the multi-band hybridization. 

Linearly polarized light will split the quadratic touching point into two Dirac points in the limit of infinite frequency.  Reducing  to finite frequency will pull the two Dirac points toward the original quadratic touching point.  At a critical frequency, the two Dirac points merge into one quadratic point. Further decreasing the frequency, the quadratic touching point will open a gap and no new Dirac points will be formed upon further lowering the frequency.  We analyze these results by deriving a 2-band effective Hamiltonian.  The results are summarized in Table~\ref{table:low_energy_gap}.  

Finally, we calculate the frequency dependent optical Hall conductivity using the four band generalized model and analyze the various inter-band contributions to the Floquet modes.   We find the smaller probe frequencies $\omega$ provide the most useful information in experiment.  In addition, a relatively small number of transitions dominate $\sigma_{xy}^{ideal}(\omega)$, which can potentially be exploited in experiments or applications that are seeking to be sensitive to a particular frequency of light and relatively insensitive to others.  This insight could be useful in searching for materials that might be good candidates for observing Floquet-Bloch states in the optical conductivity.  We hope our study will help motivate further work in this direction.

\section*{Acknowledgements} 
We acknowledge helpful discussions with Ming Xie, Qi Chen, Chao Lei and Andreas R\"uegg.  We are grateful to Xiaoting Zhou for a collaboration on a related project, and we gratefully acknowledge funding from ARO grant W911NF-14-1-0579 and NSF DMR-1507621.

%\bibliography{lno3,pyro111,Interface_DOE,kagomeref}


\end{document}